**A critical review on recent progress of solution-processed monolayer assembly of nanomaterials and applications**

*Liang Zhao, Jichao Fan, Chenchi Gong, Alexis Dyke, Weilu Gao*, Bo Li**

L. Zhao, C. Gong, A. Dyke, B. Li

Department of Mechanical Engineering, Hybrid Nano-Architectures and Advanced Manufacturing Laboratory

Villanova University

800 E. Lancaster Ave., Villanova, PA 19085, USA.

J. Fan, W. Gao

Department of Electrical and Computer Engineering

The University of Utah

201 Presidents' Cir, Salt Lake City, UT 84112, USA

L. Zhao and J. Fan contributed equally to this work.

*E-mail: weilu.gao@utah.edu; bo.li@villanova.edu

**The rapid development in nanotechnology has necessitated accurate and efficient assembly strategies for nanomaterials. Monolayer assembly of nanomaterials (MAN) represents an extreme challenge in manufacturing and is critical in understanding interactions among nanomaterials, solvents, and substrates. MAN enables highly tunable performance in electronic and photonic devices. This review summarizes the recent progress on the methods to achieve MAN and discusses important control factors. Moreover, the importance of MAN is elaborated by a broad range of applications in electronics and photonics. In the end, we outlook the opportunities as well as challenges in manufacturing and new applications.**

## 1. Introduction

Nanotechnology, with its inherent capacity to manipulate materials at the nanoscale, has truly become a game-changer in the scientific world, ushering in a host of new possibilities across numerous disciplines. At the nanoscale, materials exhibit unique properties and behaviors that are starkly different from their macroscopic counterparts. Yet, the real potential of nanotechnology is not limited to the nanomaterials' outstanding properties, but lies in our ability to assemble these minuscule components into larger, organized structures that can express and harness these unique properties in meaningful and scalable ways.[1] Among various organized structures of nanomaterials, monolayer assembly of nanomaterial (MAN) represents unique advantages in both understanding nanoscale interactions (among nanomaterials, solvents, and substrate), and designing electrical and optical properties.[2] For example, Au nanoparticle monolayer can be usually used as the substrate for surface-enhanced Raman spectroscopy (SERS) to enhance the signal of molecules.[3] Among the techniques to achieve MAN, solution-processed assembly stands out due to its precision, scalability, versatility, cost-effectiveness, and adaptability.[4] It allows for forming well-ordered monolayers of a wide range of nanomaterials over diverse substrates. This process's subtlety lies within the careful choice of assembly parameters and the understanding of forces driving the assembly, such as Van der Waals, electrostatic, and capillary interaction.[1]

This review summarizes the development of the solution-processed method for the manufacturing of MAN. Depending on the interface to form the monolayer structures, the existing methods will be summarized into three categories: air-liquid, liquid-liquid, and liquid-solid interface assembly. The fundamental principles to guide the monolayer formation and structure control (i.e., close-packed nanoparticle crystal and alignment) will be discussed. Then, diverse applications, such as photonics and electronics, are reviewed to highlight the potential of monolayer nanomaterials. Finally, we shed light on the current challenges and outlook the future directions regarding nanomaterial preparation, assembly methods, and applications (Figure 1).

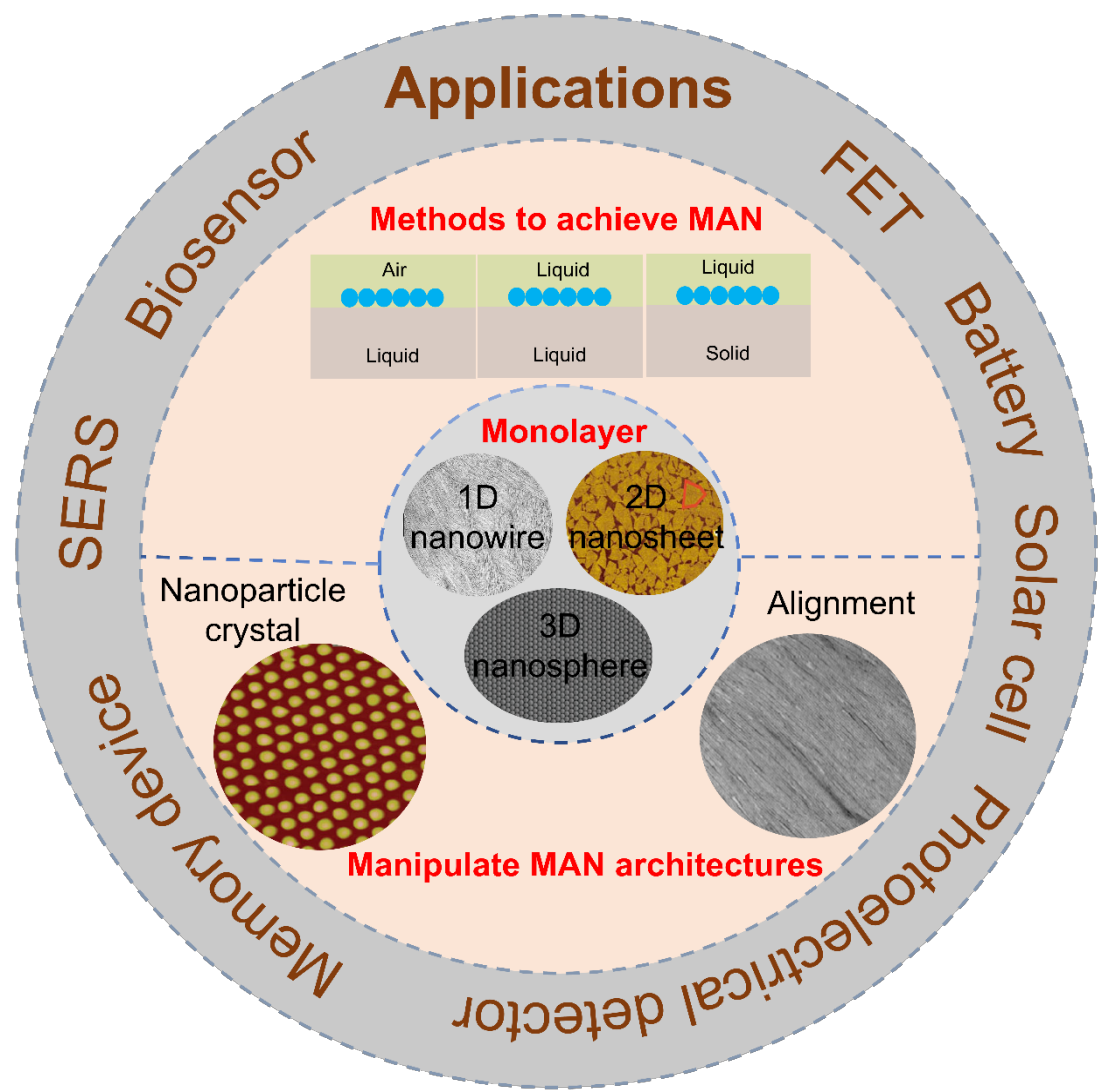

**Figure 1.** Schematic of MAN architecture and manipulation, methods, and applications. 1D nanowire figure is adapted with permission.[5] Copyright 2019, Elsevier. 2D nanosheet figure is adapted with permission.[6] Copyright 2017, American Association for the Advancement of Science. 3D nanosphere figure is adapted with permission.[7] Copyright 2022, Wiley-VCH. Nanoparticle crystal figure is adapted with permission.[8] Copyright 2015, Wiley-VCH. Alignment figure is adapted with permission.[9] Copyright 2020, Springer.

## 2. Monolayer Nanomaterial Assembly Methods

Researchers have invented different methods to achieve MAN.[10] There are different ways to classify these methods. Here, we classify these methods according to the interface where the monolayer structure forms into air-liquid, liquid-liquid, and liquid-solid assemblies. But it is also possible to divide these methods into self-limiting methods and controlled monolayer assembly methods. In self-limiting assembly methods, the self-limiting mechanism will prevent the further stacking of nanomaterials once the monolayer structure is formed. The self-limiting mechanisms include designing the attractive interaction between nanomaterials and substrates and repulsive interaction among nanomaterials, or creating interfacial confinement.[11] For example, negatively charged nanomaterials can be attracted to positively charged substrates by electrostatic attraction but repulsive interaction among nanomaterials prevents the further stacking of nanomaterials on the existing monolayer assembly. Such self-limiting characteristics enable unique fault tolerance in processing and are crucial for the scalable and cost-effective manufacture of nanomaterials monolayers.[12] On the other hand, controlled monolayer assembly methods require delicate control of processing parameters such as nanomaterial concentration and processing conditions (e.g., temperature, evaporation speed, or spinning speed of the solvent) to make sure that only one layer of nanomaterials is formed.[13] However, we do identify some overlapping cases, such as the Langmuir-Blodgett (LB) method, where the interfacial confinement at an incompatible interface (e.g., air-liquid interface) can be considered as a self-limiting mechanism. However, the number of layers will be affected by the nanomaterial concentration at the interface. Therefore, to prevent such ambiguity, we will stick to the first classification. The details discussion will be summarized in the following subsections.

### 2.1. Air-liquid Interface Assembly Methods

Air-liquid interface assembly defines a collection of methods to form MAN at the air-liquid interface. Depending on how to add nanomaterials at the air-liquid interface, there are two categories: 1) extracting nanomaterials from the liquid phase (Figure 2a), and 2) adding a nanomaterials-containing droplet as an inter-phase between the air and the supporting liquid, followed by evaporating the liquid inter-phase (Figure 2d). In both cases, eventually, nanomaterial monolayer will stably float at the air-liquid interface (Table 1). If water is used as the supporting liquid phase, for air-water interface, chemical treatments such as charge screening and hydrophobic coating are usually required to ensure the lower interfacial energy of nanomaterials at air-liquid interface. Hydrophobic polymer coatings, such as polyvinyl pyrrolidone (PVP), polystyrene (PS), polymethyl methacrylate (PMMA), and

poly(N-isopropylacrylamide) (PNIPAM), can be employed to modify the nanomaterial surface and form a shell layer.[14] The charge screening can be achieved by neutralizing negative charges, such as protons absorbed to hydroxyls on nanomaterial surface. For the two categories, the nanomaterial concentration is decisive to control the monolayer formation. In the second case, solvent for nanomaterials of the inter-phase and its evaporation speed are important for MAN formation.

In the first case, the promoter is employed to extract the nanomaterials from the liquid phase to the air-liquid interface. There are two ways that the promoter extracts the nanomaterials. First, one can add promoters to create nanomaterial vertical flow and deliver the nanomaterial to the air-liquid interface. For example, Yun et al. added ethyl acetate to the MXene aqueous solution and created a Rayleigh-Bénard convection.[14g] The convection driving force is based on endothermic evaporation of ethyl acetate to build vertical flow, bringing the MXene (surface charge screening by $HNO_3$) to the air-liquid interface. Followed by the further evaporation of ethyl acetate on water surface, the surface tension difference between ethyl acetate and water makes the MXene flow laterally and assemble into continuous monolayer film (Figure 2b). Second, the delivery process of nanomaterials to the air-liquid interface can be achieved by adding an immiscible droplet than can attach the nanomaterials. For example, Liebig et al. prepared one Au nanotriangle aqueous solution and added ethanol-toluene mixed droplets.[15] The ethanol-toluene mixed droplets can attract nanomaterials and then move to the air-water interface. Followed by evaporation of ethanol-toluene solution, monolayer structure of Au nanotriangles forms at the air-water surface (Figure 2c).

In the second case, the choice of solvent for the nanomaterials-containing droplet (the inter-phase) and solvent evaporation rate plays a crucial role in the air-liquid interface assembly. Langmuir-Blodgett method is a well-known method under this category. If water is used as the supporting liquid phase, nanomaterials will be hydrophobic or treated with hydrophobic agents (e.g., hydrophobic polymer coating or surfactant) and the nonpolar solvent (e.g., hexane, toluene, and 1-butanol) will be chosen to disperse nanomaterials uniformly.[16] Once the nanomaterials-containing droplet is placed on the air-liquid interface, the nonpolar solution will spread uniformly on the supporting liquid phase forming an extremely thin inter-phase. By controlling the concentration of nanomaterials, only one layer of nanomaterial is contained in the thin inter-phase. During the evaporation process of the inter-phase solvent, the distances among nanomaterials become smaller and smaller and eventually forms a continuous nanomaterial monolayer. For example, Han et al. coated the Au nanoparticles with a PS layer and used PS@Au nanoparticles to prepare a large-scale monolayer structure through LB method.[14h] By adjusting the solvent evaporation rate, uniform monolayer nanoparticles form on air-liquid interface after solvent evaporation (Figure 2e). There is an optimized evaporation speed of the solvent to achieve organized MAN structures such as close-packed crystal. The slower solvent evaporation cannot provide enough drive force to pull nanomaterials together to form the monolayer, and faster evaporation does not offer enough time for the reorganization of the monolayer leading to defective structures.[14h] Thus, solvent selection determines the uniformity of monolayer structure, which is also demonstrated in nanowire monolayer assembly (Figure 2f).[9]

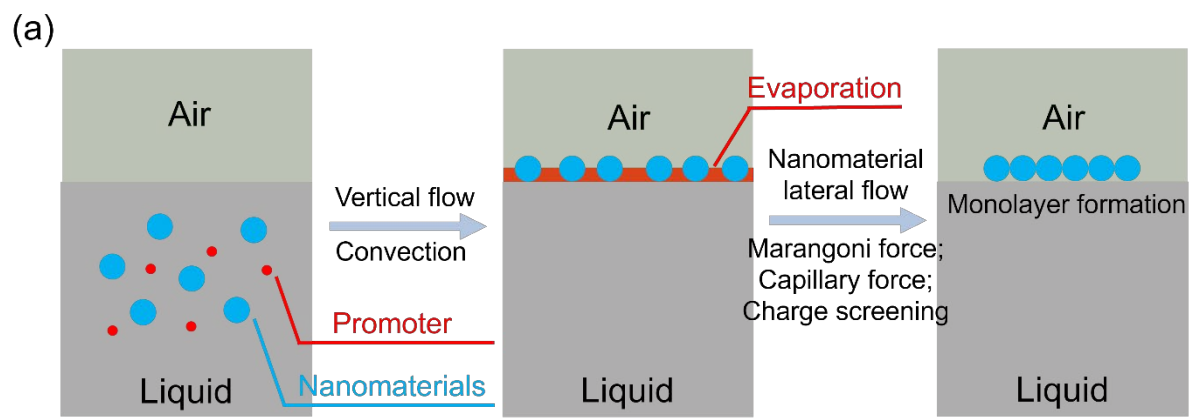

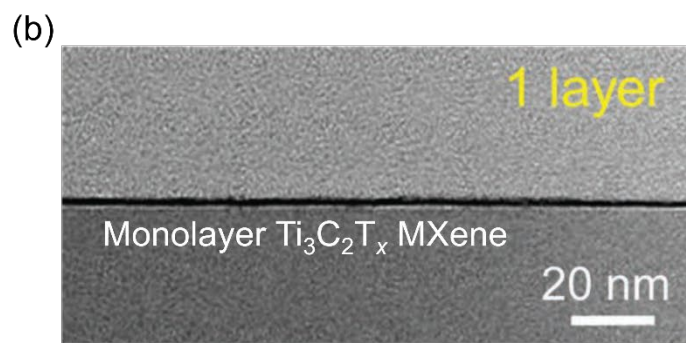
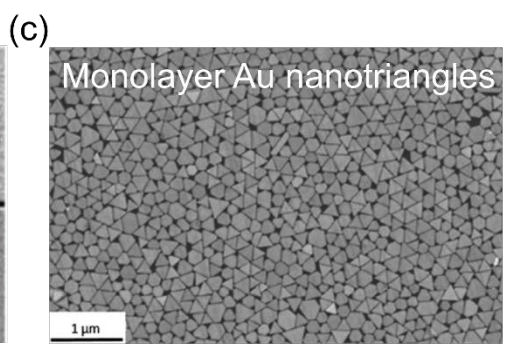

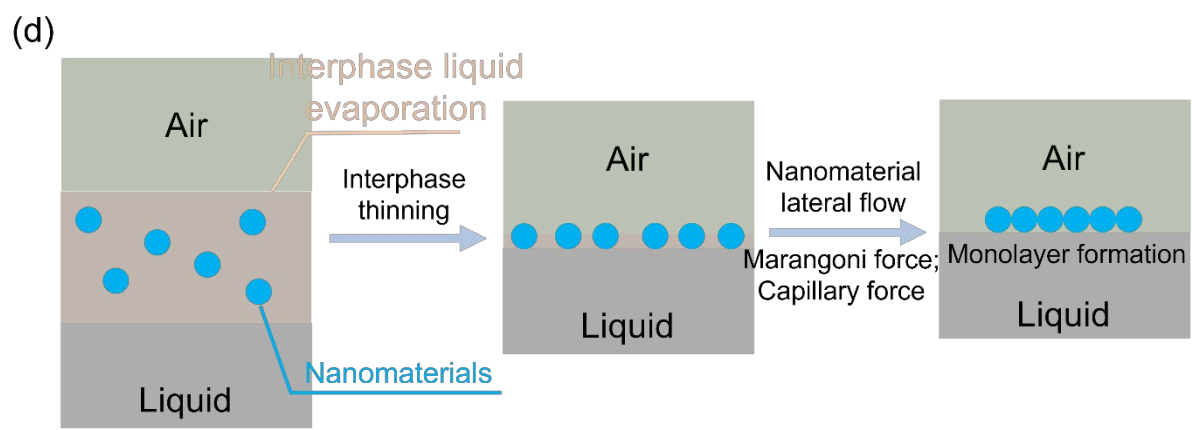

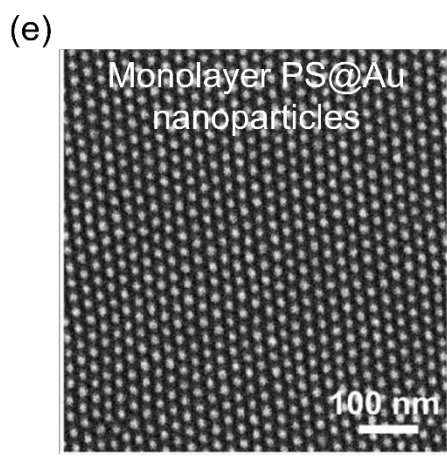
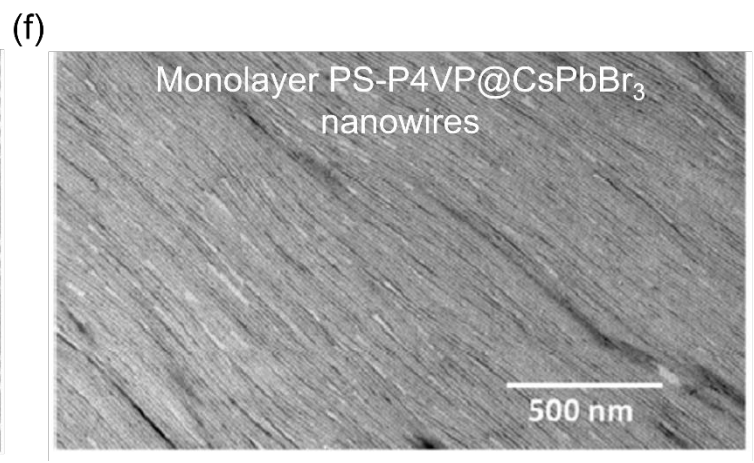

**Figure 2.** a) Schematic of air-liquid interface assembly by precipitating nanomaterials from the liquid phase. b) Cross-sectional TEM image of MXene monolayer. Reproduced with permission.[14g] Copyright 2020, Wiley-VCH. c) SEM image of self-assembled Au nanotriangle monolayer. Reproduced with permission.[15] Copyright 2017, American Chemical Society. d) Schematic of air-liquid interface assembly by adding a nanomaterials-containing droplet as an inter-phase between the air and the supporting liquid, followed by evaporating the liquid inter-phase. e) TEM images of PS@Au nanoparticle monolayer. Reproduced with permission.[14h] Copyright 2022, Elsevier. f) TEM image of PS-P4VP@CsPbBr$_3$ nanowire monolayer. PS-P4VP: polystyrene-block-poly(4-vinylpyridine). Reproduced with permission.[9]. Copyright 2020, Springer.

**Table 1.** Air-liquid interface assembly of nanomaterial monolayer.

| Nanomaterials | Modification | Size (nm) | Solvent | Concentration | Liquid subphase | Refs. |
|---|---|---|---|---|---|---|
| Au and $Fe_3O_4$ NP | Au: dodecanethiol $Fe_3O_4$: oleate | Au: 5.2 $Fe_3O_4$: 10.6 | Hexane | 0.1-1 mg/mL (mixture ratio:1:1) | Diethylene glycol | [16] |
| $CsPbBr_3$ NW | PS-P4VP coating | Diameter: 10 Length: N/A | Toluene/hexanes: 1/1, v/v | N/A | Acetonitrile (55 vol%)/water | [9] |
| $Ti_3C_2T_x$ nanosheet | N/A | Lateral size: 5000 Thickness: 1.3 | $HNO_3$ (1 M)/water | 2.9 ug/mL | $HNO_3$ (1 M)/water | [14g] |
| ZIF-8 MOF NP | PMMA coating | 174 | Toluene | 30 mg/mL | Water | [14e] |
| Te NW | PVP | Diameter: 10 Length: Several micrometers | 1-butanol | N/A | Water | [14a] |
| Au nanoparticle | PS coating | 8, 16, 24 | $CHCl_3$ | 60-200 mg/mL | Water | [14d] |
| Ag nanorod | PVP | Diameter: 32 Length: 92, 204, 383, 515, 771, 910 | Ethanol/dichloromethane/cyclohexane/octane: 1:1:1:0.5, v/v/v/v | N/A | Water | [14b] |
| Au@Ag nanoparticle | PNIPAM | 326 | Ethanol | 2 wt% | Water | [17] |
| ZIF-8 MOF nanoparticle | DPGG | 552.3 | Toluene | 30 mg/mL | Water | [14f] |
| CuO nanosheet | N/A | Lateral size: 30-50 Thickness: 300-1000 | Ethanol | 0.2 mg/mL | Water | [18] |
| $SiO_2$ microsphere | N/A | 951 | Ethanol | 0.01-0.2 vol% | Water | [19] |
| Au nanoparticle | PVP | 45 | Water | N/A | Water | [14c] |
| Au nanotriangle | N/A | Edge length: 175 | Ethanol/toluene | $3.5 \times 10^{-4}$ g/cm$^3$ | Water | [15] |
| CdSe nanoplatelet | N/A | N/A | Hexane, orctane | N/A | Acetonitrile | [20] |
| F,Sn:$In_2O_3$ nanocube | N/A | Edge length: 12.6 | Hexane | 1 mg/mL | Ethylene glycol | [21] |
| PbSe | oleic acid | 4.5 and 6.0 | Hexane | 10 mg/mL | diethylene glycol | [22] |
| Au nanoparticle | N/A | 35 | Water | N/A | Water | [13a] |
| Au nanoparticles | PS | 15.3 | Chloroform/toluene/mesitylene: 2:1:1, v/v/v | 1 mg/mL | Water or ethylene glycol | [14h] |
| Au nanoparticles | PVP | 50 | Ethanol | N/A | Water | [23] |
| $TiO_2$ nanoparticle | Oleic acid and oleylamine | Longitudinal: 5.7 Transversal: 4.6 | Pentane/dichloromethane/oleylamine/ hexane | pentane/dichloromethane 3:1 | Water | [24] |
| Single-wall carbon nanotube | Poly(p-phenylenevinylene-co-2,5-dioctyloxy-m-phenylenevinylene | Diameter: 1.4 | 1,2-dichloroethane | N/A | Water | [25] |

## 2.2. Liquid-liquid Interface Assembly Methods

The liquid-liquid interface assembly methods achieve highly reproducible and defect-free for the assembly of nanomaterial monolayer at the interfaces of two immiscible liquid phases (e.g., the oil and water, Table 2).[26] The difference for the surface tensions of two liquids will generate an interface that has high interfacial tension. If adding nanomaterials to such high energy interface can reduce the total potential energy of the system, the nanomaterials can be trapped at the interface. By carefully tailoring nanomaterial concentration, a MAN can be formed at liquid-liquid interface.[26]

Basically, there are two ways to enable the liquid-liquid interface assembly of nanomaterial monolayer. First, nanomaterials can be extracted from one of the liquid phases and form an energetically stable monolayer at the liquid-liquid interface. For example, Song et al. mixed Au nanoparticle aqueous solution with perfluorodecanethiol (PFT)/ethanol/hexane solution (Figure 3a).[13a] The PET molecule can be attached on Au nanoparticle surface and lower its surface energy, and the PET@Au nanoparticles tends to quickly achieve monolayer structure at the high interfacial tension between ethanol/hexane and water to reduce the system potential energy (Figure 3b). Besides, this liquid-liquid interface assembly method can also be applied to a wide range of nanomaterials with different shapes such as Au nanorods, Ag nanoparticles, and $Fe_3O_4$ nanoparticles (Figure 3c).[13a] Second, the nanomaterial solution can be directly injected into the liquid-liquid interface and form monolayer. For example, Yu et al injected $WSe_2$/hexylamine solution at the interface between hexane and ethylene glycol (Figure 3d).[27] $WSe_2$ is functionalized by hexyl trichlorosilane. As the hexylamine rapidly dilutes into the hexane layer, the functionalized $WSe_2$ nanosheets are confined by interface with 2D flakes aligned parallel to the 2D interface. Increasing the loading of flakes leads to in-plane self-compression and generates a compact 2D monolayer of the $WSe_2$. Next, the hexane layer is removed from the top leaving the $WSe_2$ monolayer floating on the surface of the ethylene glycol (Figure 3e-3g).

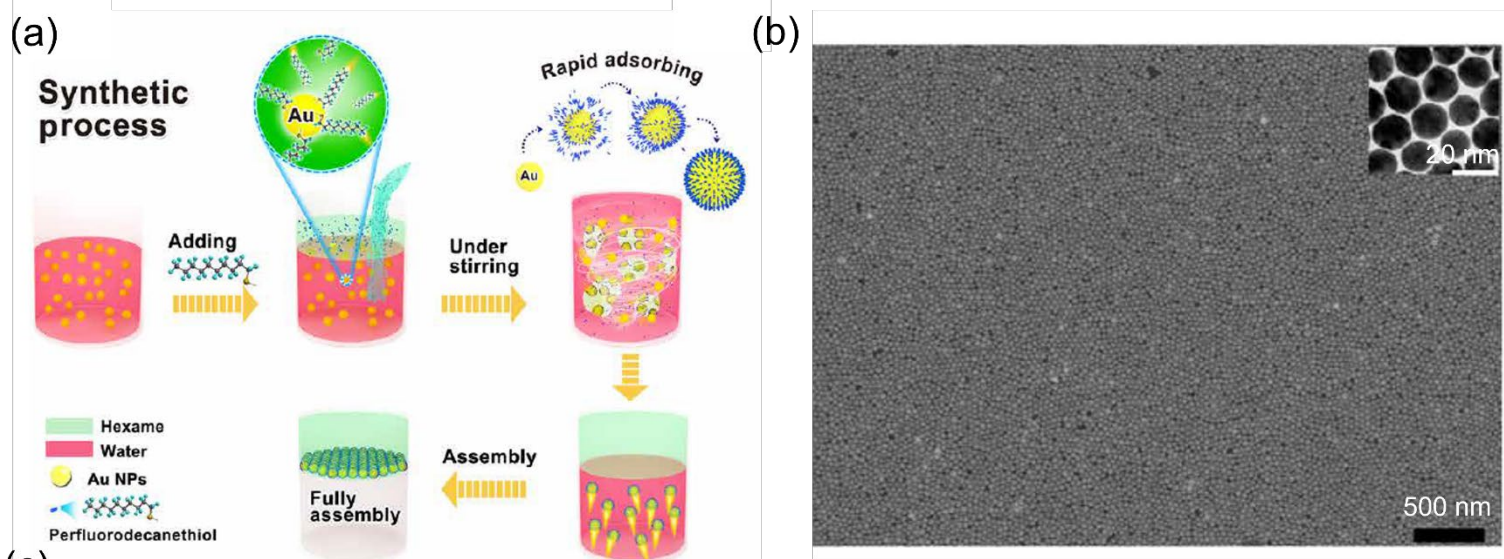
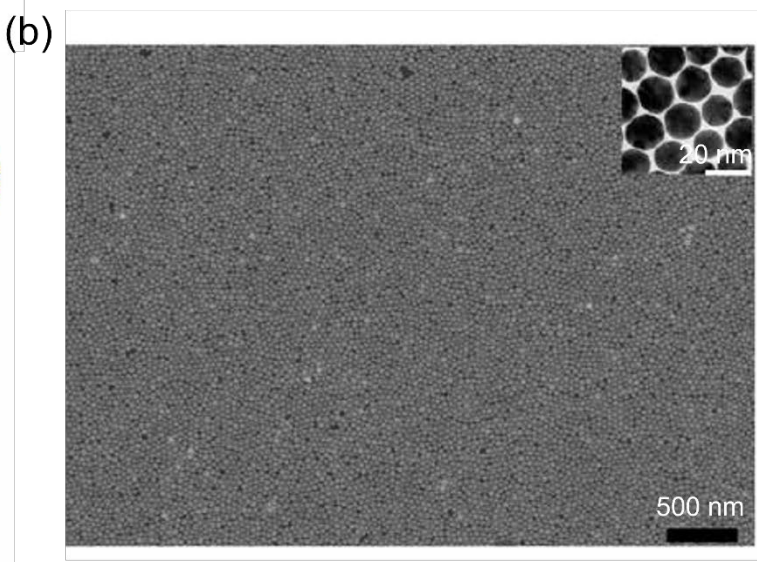
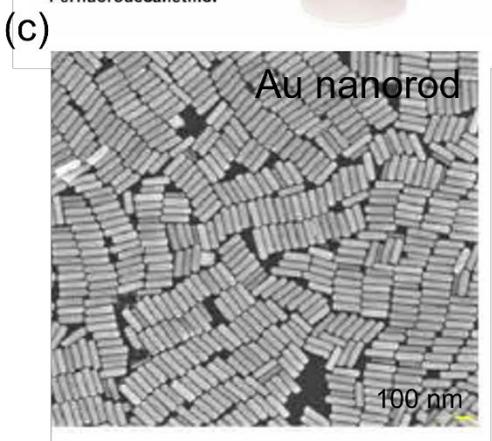
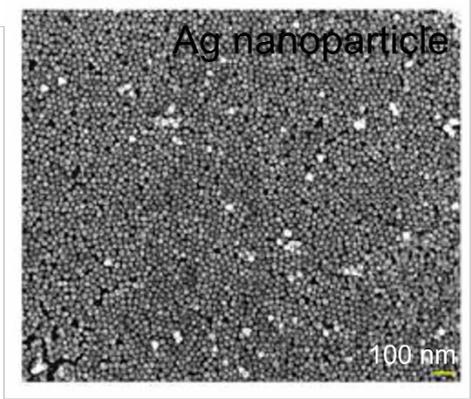
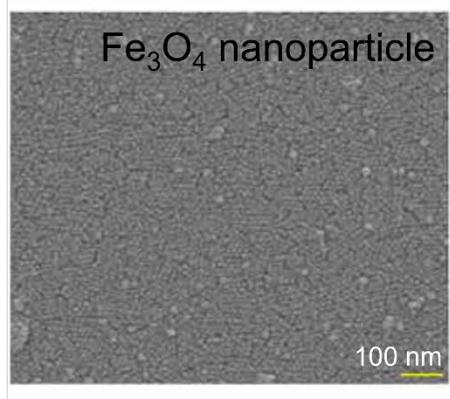
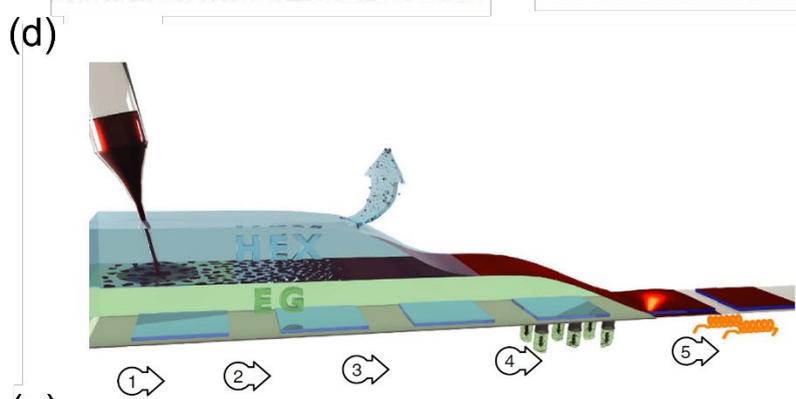
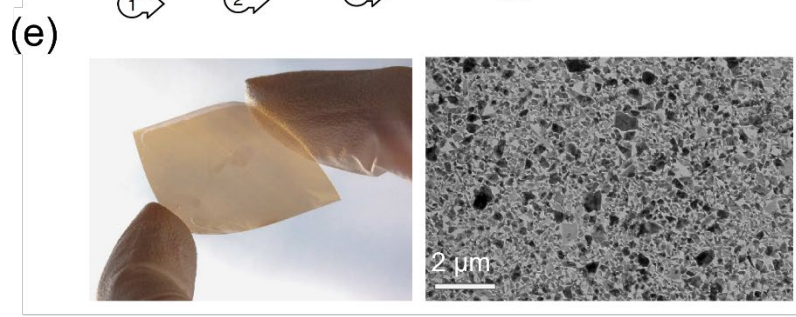
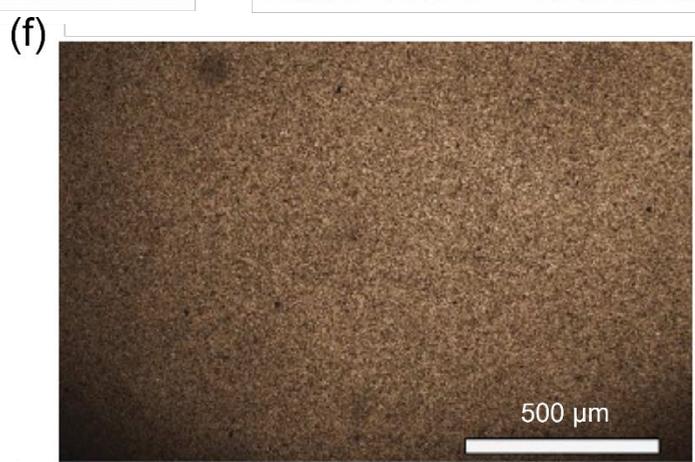
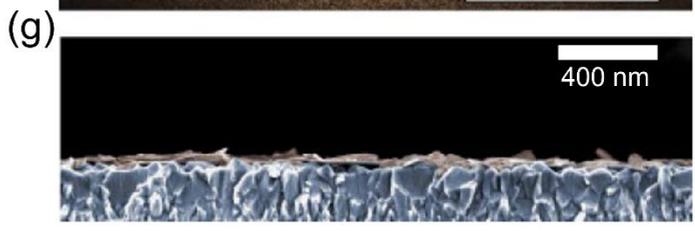

**Figure 3.** a) Schematic illustration of liquid-liquid interface assembly of monolayer nanomaterials induced by perfluorodecanethiol (PFT). b) SEM and TEM (inset) images of monolayer Au nanoparticle monolayer. c) SEM images of monolayer assembly of Au nanorods, Ag nanoparticles, and $Fe_3O_4$ nanoparticles. Adapted with permission.[13a] Copyright 2021, American Association for the Advancement of Science. d) Schematic of monolayer $WSe_2$ nanosheet assembly at liquid-liquid interface. Step 1: injection of $WSe_2$ dispersion; Step 2: flake confinement and self-assembly; Step 3: hexane removal; step 4: ethylene glycol removal and monolayer film deposition; step 5: drying at 150 °C. e) A photograph of a monolayer $WSe_2$ thin film on ITO-coated PET film and TEM image of a monolayer $WSe_2$. f) optical image of monolayer $WSe_2$ film and g) corresponding cross-sectional SEM image. Adapted with permission.[27] Copyright 2015, Nature Publication Group.

**Table 2.** Liquid-liquid interface assembly of nanomaterials

| Nanomaterials | Surface chemistry | Size (nm) | Solvent | Concentration (mg/mL) | Liquid-liquid | Promoter | Ref |
|---|---|---|---|---|---|---|---|
| Au nanoparticle | PVP | 32-72 | Ethanol/$CH_2Cl_2$/water | N/A | Ethanol/$CH_2Cl_2$-water | n-hexane | [28] |
| Au-Ag nanorod | N/A | 91.02 in length 28.51 in diameter | Water | N/A | Cyclohexanewater | Ethanol | [29] |
| PbSe nanoparticle | N/A | 7.3 | Hexane | 0.001 | ethylene glycol | N/A | [30] |
| $Fe_2O_3$ nanoparticle | Oleate | 8.6, 11.8 | Heptane | N/A | Diethylene glycol-heptane | N/A | [31] |
| Au nanoparticle | | 60 | | | Heptane-1,2-dichloroethane/water | | |
| $WSe_2$ nanosheet | Hexyl-trichlorosilane | Thickness: 10 | Hexylamine | 10 | Ethylene glycol-hexane | N/A | [27] |
| PS-PB-PMMA nanoparticle | N/A | 55 | Toluene | 8 | Tuluene-LiCl/water | N/A | [32] |
| Au nanoparticle | Citrate | N/A | Cetyltrimethylammonium bromide/water | $5 \times 10^{11}$ particles $mL^{-1}$. | Dichloromethane-cetyltrimethylammonium bromide/water | Cetyltrimethylammonium bromide | [33] |
| $MoS_2$ nanosheet | N/A | Thickness: 2 | IPA | N/A | Water-hexane | | [34] |
| Au nanoparticle | PVP coating | 28.8 | Ethanol/$CH_2Cl_2$/water | N/A | Ethanol/$CH_2Cl_2$/water -n-hexane | | [35] |
| Au nanorod | PEG-SH | N/A | Water | N/A | Water/Hexane | Ethanol | [36] |
| Pd@Au nanoparticle | N/A | 15, 16.9, 18.4 | Water | N/A | Hexane/Water | Ethanol | [37] |
| Carbon nanotube | Poly[(9,9-dioctylfluorenyl-2,7-diyl)-alt-co-(6,6′-{2,2′-bipyridine})] | Diameter: 1.5 | Chloroform | N/A | Chloroform/Water | N/A | [38] |
| Carbon nanotube | Poly [9-(1-octylonoyl)-9H-carbazole-2,7-diyl] | Diameter: 1.51 | 1, 1, 2-trichloroethane | N/A | 1, 1, 2-trichloroethane/2-butene-1,4-diol | N/A | [39] |

## 2.3. Liquid-solid Interface Assembly Methods

Assembly of a nanomaterial monolayer at the liquid-solid interface usually relies on precise control of the adsorption and arrangement of nanoparticles at the interface. To achieve such an assembly process, multiple methods can be utilized such as immersing, drop casting, dip coating, spray coating, and spin coating (Table 3). The mechanisms behind those methods can be divided into three categories: forced deposition, chemical interaction-based assembly, and physical interaction-based assembly.

Forced deposition requires delicate processing parameter manipulations. Evaporation-driven monolayer assembly is a typical forced deposition method (Figure 4a). During the evaporation process, nanomaterials are forced to precipitate from the oversaturated wetting solution on the surface of the substrate. This mechanism can be integrated with multiple manufacturing processes such as dip coating and spin coating to achieve monolayer assembly. Taking dip coating as an example, the nanomaterials containing solution (e.g., water) wet up the substrate forming a thin film (Figure 4b).[13b] By controlling the concentration of nanomaterial in the wetting film, a MAN can be formed during solvent evaporation. For example, Núñez et al. achieved monolayer $SiO_2$ nanoparticles on Si wafer using dip-coating method.[13b] The challenge of evaporation-driven assembly is the "coffee ring" effect caused by contact line pinning during evaporation where nanoparticles move toward the pinned contact line and leave a coffee ring.[40] To solve the problem and achieve a uniform monolayer, Zargartalebi et al. developed a meniscus-free evaporation process to deposit monolayer particles onto a superhydrophilic substrate.[41] The superhydrophilic substrate is set as the base and is surrounded by vertical walls that have a near-neutral wettability. In this way, a flat liquid-air interface is formed without any bending near the vertical wall (e.g., no wet up or bend down), which eliminates the capillary and Marangoni flow towards the wall and eventually forms a monolayer structure on the base (Figure 4c).

Chemical interaction-based assembly builds effective interaction between solid substrates and nanomaterials (Figure 4d-e). Such interactions include electrostatic attraction, DNA matching, and ligand exchange. For example, Zhang et al. assembled positively charged Au nanoparticles (treated with poly dimethyl-diallyl ammonium chloride) on a negatively charged ITO-coated glass substrate (Figure 4f).[11] Because of the electrostatic attraction between Au nanoparticles and ITO-glass substrate and electrostatic repulsion between Au nanoparticles, a monolayer structure can be achieved (Figure 4g). Another method is matchable DNA modifications on both nanoparticles and substrate. The Au nanoparticles and substrates are both decorated with oligonucleotide, through a simple immersing and shaking process, an Au nanoparticle monolayer can be achieved at specific locations. The post-washing process ensures monolayer formation, avoiding excessive nanomaterials remaining, where the unmatchable DNA chains among Au nanoparticles also avoid second layer deposition. The third method is based on ligand exchange strategy (Figure 4h). The hydrogen terminated or undecenethiolate modified boron doped diamond substrates experience a ligand exchange process with matchable ligand grafted Au nanoparticle (Figure 4i).[42] The attractive interaction between nanomaterials and substrates is necessary for the formation of monolayer assembly but not enough. The surface chemistry should be designed to minimize the attraction of nanoparticles or even create repulsion mechanisms to prevent the aggregation of nanoparticles in the solvent and the assembly of the second layer. Moreover, post-assembly rinsing of the assembled substrates is also important, because excessive containing nanomaterials solution on the substrate may deposit add-on layers through the evaporation process.

The physical interaction-based assembly relies on van der Waals interactions among, nanomaterials, solvents, and substrate.[12a] The key is the designed an energetically favorable scenario for the assembly of nanomaterials. For example, Zhao et al., developed an acoustic self-limiting assembly method to achieve monolayer $SiO_2$ nanoparticle on a polydimethylsiloxane (PDMS) in water (Figure 4j).[12a] $SiO_2$ nanoparticle is treated with (3-aminopropyl) triethoxysilane to convert its hydrophilic surface into hydrophobic surface. The hydrophobic $SiO_2$ nanoparticle tends to stabilize hydrophobic PDMS substrate in water. The self-limiting mechanism here is based on physical collision. The acoustic field energizes the $SiO_2$ nanoparticles to collide with PDMS substrate. The PDMS is a viscoelastic elastomer that can absorb the collision energy of nanoparticles and capture the nanoparticles. However, after the formation of the first layer, the following nanoparticles will collide with elastic $SiO_2$ nanoparticles and be bounded off (Figure 4k). This unique mechanism enables an eco-friendly, ultrafast assembly (2 min to achieve full coverage) (Figure 4l), and scalable method to achieve MAN.

It is also important to note that some mechanisms can be combined. For example, Matsuba et al. used spin coating method to assemble 2D materials on Si substrate in the form of monolayer films.[6] The positively charged 2D nanosheets dispersed in DMSO solvent repel each other due to electrostatic repulsion, which avoids aggregation and effective interaction between nanosheets and substrates (Figure 4m). Also, the spinning process thins the solution on the Si substrate and induces a uniform evaporation process (i.e., forced deposition), which ensures the monolayer structure (Figure 4n).

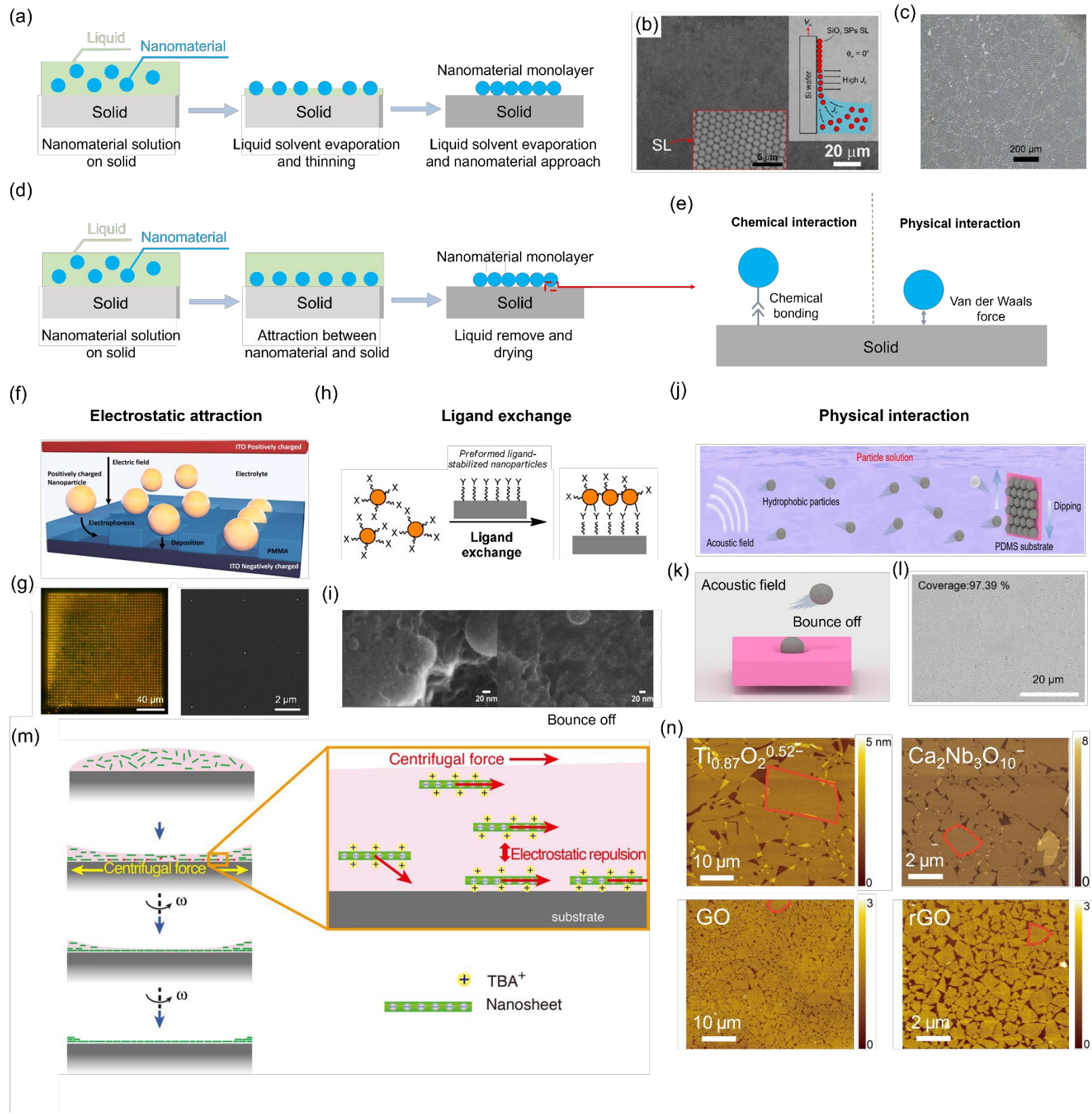

**Figure 4.** a) Schematic of forced deposition assembly of nanomaterial monolayer. b) SEM images of SiO$_2$ nanoparticle monolayer prepared by dip-coating method. The inset is schematic of the dip-coating conditions. Adapted with permission.[13b] Copyright 2018, American Chemical Society. c) SEM image of PS particle monolayer. Adapted with permission.[41] Copyright 2022, Nature Publication Group. d) Schematic of chemical and physical interaction-based assembly of nanomaterial monolayer and e) corresponding interaction between nanomaterial and solid substrate. f) Schematic of the electrophoretic deposition process showing positively charged nanoparticle deposited onto a negatively charged ITO substrate. g) Dark-field image and localized SEM image of an Au nanoparticle array. Reproduced with permission.[11] Copyright 2018, American Chemical Society. h) Schematic of monolayer nanoparticles assembled on a substrate through ligand exchange. i) SEM images of undecenethiolate-Au nanoparticle monolayer assembled on a hydrogen terminated boron doped diamond substrate and triphenylphosphine-Au nanoparticle monolayer assembled on a undecenethiolate modified boron doped diamond substrate. Adapted with permission.[42] Copyright 2016, American Chemical Society. j) Schematic of the acoustic field assisted assembly of monolayer structure based on physical interaction. k) Schematic of second nanoparticle bouncing off due to the acoustic field. l) SEM image of hydrophobic SiO$_2$ nanoparticle monolayer assembled on PDMS substrate. Reproduced with permission.[12a] Copyright 2022, American Chemical Society. m) Assembly of edge-to-edge monolayer nanosheets using spin-coating process. n) AFM images of Ti$_{0.87}$O$_2^{0.52-}$, Ca$_2$Nb$_3$O$_{10}^-$, GO, and rGO monolayer nanosheets. Adapted with permission.[6] Copyright 2017, American Association for the Advancement of Science.

**Table 3.** Liquid-solid interface assembly of nanomaterials.

| Nanomaterials | Modification | Size (nm) | Solvent | Concentration (mg/mL) | Solid substrate | Substrate modification | Assembly method | Ref |
|---|---|---|---|---|---|---|---|---|
| Au nanoparticle | α-Synuclein | 10<br>20<br>30<br>100; | Citrate buffer, | N/A | Polycarbonate | Oxygen plasma | Drop casting | [43] |
| Au nanoparticle<br>Au nanorod | poly-DADMAC | 100 in length and 40 in width | NaCl aqueous solution | 0.03 | ITO-glass | N/A | Electrophoretic deposition | [11] |
| Au nanoparticle<br>Ag nanoparticle | Citrate | 45; 49 | Toluene | 1 | AAO template | N/A | Sonication | [44] |
| Ag nanowire | PVP | 47 in width 4200 in length | Water | 0.12 | Glass slide or silicon wafer | PEI coating | Spray coating | [45] |
| Au nanoparticle | PS | 49.4<br>76.3<br>120.5 | Water | N/A | Silicon wafer | N/A | Drop casting | [46] |
| Au nanorod | CTAB | 40 in width<br>110 in length | N/A | N/A | Si wafer | N/A | N/A | [47] |
| Au nanosphere<br>Au nanocube<br>Au triangular prism | Oligonucleotide | 60-100 | Water | N/A | Gold | Oligonucleotide | Immersing | [12c] |
| Au nanoparticle | Citrate | 10-12 | Water | N/A | PS-b-P4VP | N/A | Immersing | [48] |
| SiO$_2$ nanoparticle | stearyl alcohol | 50 | toluene | 0.1 vol% | Si wafer | N/A | Drop casting | [49] |
| Au nanoparticle | Triphenylphosphine | N/A | Tetrahydrofuran | 0.1 | Boron doped diamond | S-10-(undecenyl) thioacetate | Immersing | [42] |
| SiO$_2$ nanoparticle | N/A | N/A | Water | 0.05 wt% | Glass slide | (3-aminopropyl) trimethoxysilane | Immersing | [50] |
| SiO$_2$ nanoparticle | N/A | 200 | Dimethyl Formaldehyde | 50 | Si wafer | NH$_4$OH: H$_2$O$_2$:H$_2$O | Spin coating | [51] |
| MnO$_2$ nanoparticle | Oleic acid | 30 | hexane | N/A | Pt | N/A | Dip coating | [52] |
| GO, rGO, Ti$_{0.87}$O$_2^{0.52-}$, Ca$_2$Nb$_3$O$^{10-}$ nanosheet | Tetrabutylammonium hydroxide | 2.3, 1.1, 1.4, 1 in thickness; 2000, 1000, 10000, 2000 in lateral size | Dimethylsulfoxide | N/A | Si wafer | H$_2$SO$_4$ | Spin coating | [6] |
| SiO$_2$ nanoparticles | (3-aminopropyl)triethoxysilane | 487, 761 | Water | 10 mg/mL | PDMS | N/A | Dip-coating | [12a] |
| Single-wall carbon nanotubes | Sodium dodecyl sulfate | Diameter: 0.73, 1.5 | Water | 1 - 5 μg/mL | Si wafer | N/A | Immersing | [53] |

## 2.4. Summary

In this section, we have summarized the methods to achieve MAN structures according to their formation interfaces: air-liquid, liquid-liquid, and liquid-solid interface. A collection of methods such as LB method, dip coating, and spin coating can be classified under these categories. However, these methods are also used to achieve multilayer assemblies. To guarantee MAN formation, innovative design and deliberate control are necessary. First, it is necessary to design an energetically preferred scenario to stabilize nanomaterial at these interfaces, which can be done by deliberately choosing the nanomaterial, inter-phase solvent, liquid phase, and solid phase, and surface modifications through adding promotors and chemical groups. Second, it is critical to limit the number of nanomaterials at the interface such that only one-layer forms. This requires delicate control of the processing parameters such as nanomaterial concentration and evaporation speed of the solvent. It is important to note that there are different ways to classify these methods with different emphasis of the MAN formation mechanism. For example, some methods contain self-limiting mechanisms where the assembly will terminate automatically after the formation of the monolayer, while some methods do not have it and therefore require delicate control of the processing parameters. Finally, formation of MAN is not the end of the story, and the architecture of the MAN can be further controlled. For example, the nanomaterials in the monolayer can be randomly arranged or form ordered structures. These ordered structures can be extremely important to the application of MAN. Our discussion will continue in the following two sections.

## 3. Tunable Monolayer Architectures

Most of the monolayer assembly methods discussed in the previous sections are generic towards nanomaterials with different sizes and shapes. When it comes to the architectures of these assemblies, we have identified clear size and shape dependency. For example, 3D nanoparticles and 1D nanomaterials (e.g., nanorods, nanotubes, nanowires) with narrow size distribution usually form ordered structures such as superlattice and liquid crystal phase, respectively. However, 2D flakes usually form random network because of the wide size distribution of these flakes. The ordered architectures offer desired functionalities in electronics and photonics. In this section, we will summarize methods for controlling the monolayer architectures.

## 3.1. Monolayer Nanoparticle Crystals

A monolayer nanoparticle crystal is defined as a periodic structure of a layer of nanoparticles. Here, the nanoparticle refers to 3D particles that have narrowly distributed shapes (e.g., sphere, cube, star, polyhedron, and spindle) and size. Each nanoparticle is situated at the nodal points of the crystal lattice, creating a highly ordered and periodic arrangement. The lattice may exhibit a specific symmetry, such as tetragonal or hexagonal, depending on the spatial organization of these nanomaterials.[14d, 21, 54] This arrangement is significant in materials science for understanding the mechanical, electrical, and thermal properties of monolayer nanomaterials.[43, 55] The analysis of such structures often involves lattice constants, unit cells, and the interactions between neighboring nanomaterials, which determines the macroscopic behavior of the monolayer architectures.

### 3.1.1. Close-packed Monolayer Nanoparticle Crystals

Close-packed monolayer nanoparticle crystal represents a significant advancement in nanomaterial engineering, offering unparalleled precision and efficiency in various applications such as the hexagonal structure for structural coloration,[12a] enhanced electrical performance, and increased mechanical performance.[56]

The nanosized spheres, cubes, and polyhedrons with narrow size distribution have shown successful examples of monolayer nanoparticle crystal. It is important to note these nanoparticle crystals range from perfect superlattice (like a single crystal without any defects, Figure 5a) to polycrystal with various defects (Figure 5b-f). The perfectness of the resultant crystal is determined by the narrow size distribution of nanoparticles as well as the subtle balance of nanoparticle-solvent-substrate interactions during the processing. The first landing locations of nanoparticles at interfaces are usually random, to achieve close-packed structure, extra driving force(s) is needed. For air-liquid interface assembly, the solvent evaporation between nanoparticles will exert meniscus force to pull the particles together. As we have discussed previously, solvent evaporation speed critically determines the reorganization kinetics of the particles and thereafter the crystal structure. Similarly, the liquid-liquid interface assembly also depends on the nanoparticle-nanoparticle interaction and meniscus formation. When the nanoparticle-containing solvent diffuses to one of the liquids, meniscus force caused by reduced liquid among nanoparticles tends to pull the nanoparticles together. For liquid-solid interface assembly, there is a competition between nanoparticle-solid interaction and nanoparticle-nanoparticle interaction as well as nanoparticle-liquid meniscus force. If nanoparticle-solid interaction is too strong (e.g., with strong chemical bonding), the nanoparticles will be held at the original locations on the solid substrate (usually in a random distribution) and cannot reorganize into a close-packed crystal. In the case of forced deposition, the solvent evaporation speed determines the perfectness level of the crystal. For the physical interaction-based assembly, there is no meniscus formation and the nanoparticle reorganization is driven by shear field from the moving substrate and the kinetic energy of the particle hitting on the monolayer.

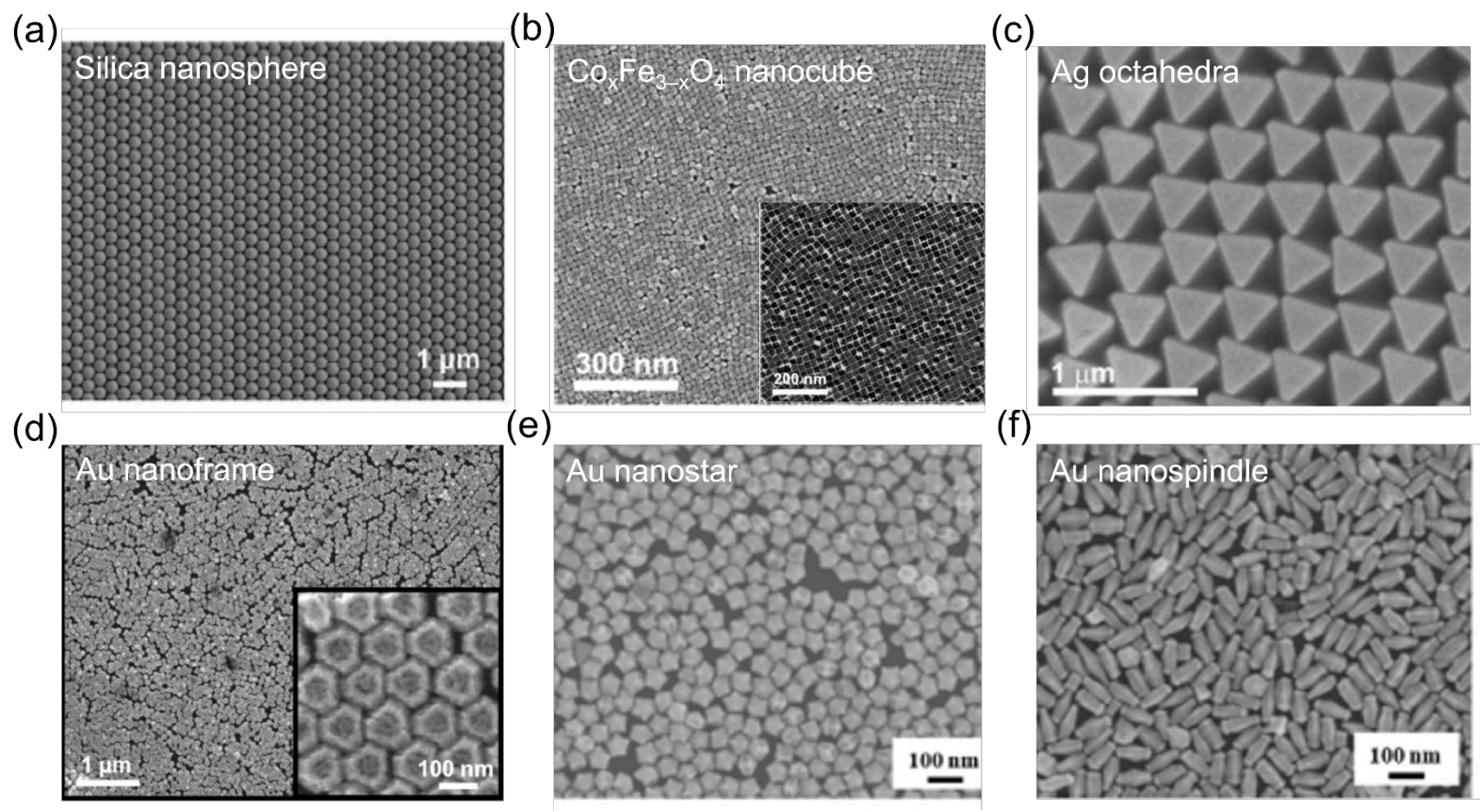

**Figure 5.** Close-packed nanoparticle monolayer structure. SEM images of a) Silica nanosphere. Adapted with permission.[7] Copyright 2022, Wiley-VCH. b) $Co_xFe_{3-x}O_4$ nanocube. Inset is a representative TEM image. Adapted with permission.[57] Copyright 2014, American Chemical Society. c) Ag octahedra. Adapted with permission.[58] Copyright 2018, Nature Publication Group. d) Au nanoframe. Adapted with permission.[59] Copyright 2022, American Chemical Society. e) Au nanostar. f) Au nanospindle. Adapted with permission.[35] Copyright 2020, Elsevier.

*3.1.2. Monolayer Nanoparticle Crystals with Interparticle Gap*

Monolayer nanoparticle crystals with controlled interparticle gaps attracted great interest because of special and tunable light-matter and electron-matter interactions.[14d, 60] Take the monolayer plasmonic nanoparticles for example, the small interparticle nanogaps can generate intense localized electromagnetic fields, which can be used to detect molecules even with a low concentration.[61] Therefore, controlling the gaps among nanoparticles in monolayer structure is of great significance in developing extreme sensing capabilities and quantum manipulation. The methods can be classified into two categories: (1) steric hinderance methods, and (2) template methods.

Basically, a general method to regulate the gaps is to design the steric hinderance (Figure. 6a). For example, Wang et al. developed polystyrene (PS) polymer ligand grafted Au nanoparticles and assembled these grafted Au nanoparticle monolayer by liquid-liquid interface assembly.[14d] The gap among Au nanoparticles is determined by the thickness of PS shell. As increasing the molecular weight of PS, the interparticle gaps appeared to be larger with well-defined hexagonal structure (Figure. 6b). This obvious gap variation is ascribed to the steric repulsion enhanced by the larger PS chains. Besides, the research also synthesized shells that can be stimulated by the environmental conditions (e.g., water and UV light), resulting in tunable nanoparticle gaps.[8, 62] For example, Volk et al. developed a silver@gold-poly-N-isopropylacrylamide (Ag@Au-PNIPAM) core-shell particles that can experience a swelled behavior when it is in wetted state (hydrogel shell).[8] With extending dwelling time, the interparticle gaps increase because of the swollen hydrogel. After drying, the interparticle distance can be extended and form monolayer with periodic gaps (Figure. 6c). Assembling a binary monolayer following etching one is also an efficient way to manipulate the monolayer nanoparticle gaps, such as $Fe_3O_4$ and Au binary nanoparticle monolayer. Once the binary nanoparticle monolayer forms, e-beam exposure enables removement of either one of the nanoparticles and convert it to uniform interparticle gap mediated monolayer structure (Figure. 6d-e).[16]

Another strategy is based on the templated substrates, which usually requires patterning techniques (e.g. photolithography, e-beam lithography, and aluminum oxide (AAO) template) to create arrays of "holes" to attract nanoparticles to be assembled on targeted locations. For examples, Udayabhaskararao et. al. created arrays of holes using e-beam lithography, and then decorated the holes with DNA (Figure. 6f).[12b] The DNA-mediated gold cubes can identify and land on the holes (Figure. 6g). It should be noting that even though the topological confinement can guide the nanoparticle deposition, external force, such as chemical bond and coulomb interaction,[11, 12b] would be applied to ensure effective nanoparticle-substrate interaction, and thereby forming less defective monolayer structure with controllable gaps. Another example is using AAO template to guide the deposition of Au nanoparticles (Figure. 6h).[44] The Au nanoparticles can be inserted into matchable holes by ultrasonication force to monolayer structure with tailored interparticle gaps.

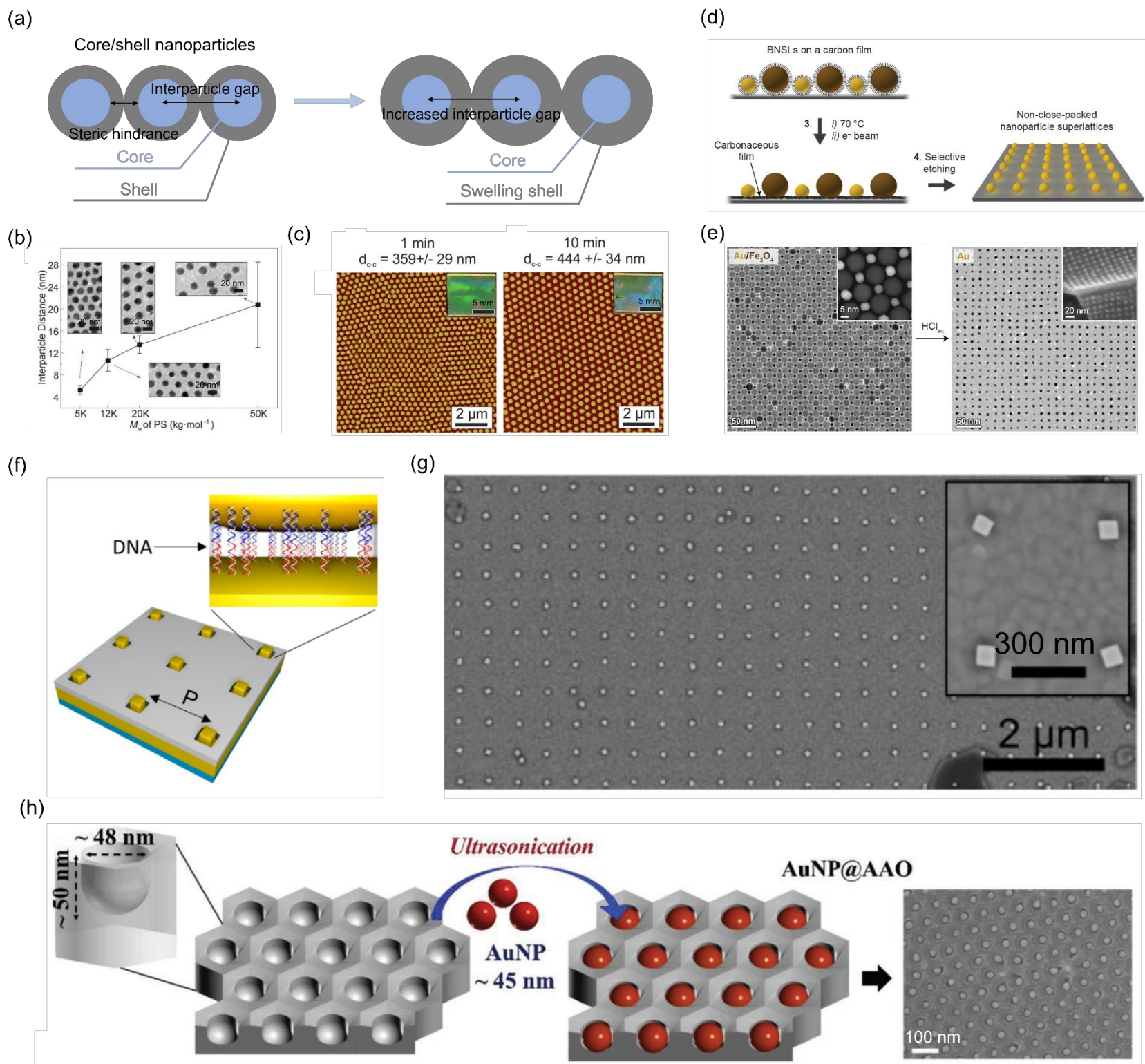

**Figure 6.** a) Schematic of steric hindrance induced gaps by core-shell structure of nanoparticles. b) PS shell molecular weight ($M_w$)-dependent interparticle distance of monolayer of PS coated Au nanoparticles (nanoparticle size: 16 nm).

Insets are the representative TEM images of different interparticle distance. Adapted with permission.[14d] Copyright 2018, Wiley-VCH. c) Atomic force microscopy (AFM) images of silver@gold-poly-N-isopropylacrylamide (Ag@Au-PNIPAM) monolayers on glass substrate withdrawn after 1 min and 10 min dwell time. The insets show digital photographs of the monolayers. Adapted with permission.[8] Copyright 2015, Wiley-VCH. d) Schematic illustration of Au and $Fe_3O_4$ binary nanoparticle monolayer and selective etching process. e) TEM image of a binary monolayer superlattice of $Fe_3O_4$ and Au nanoparticles and SEM image of selective removal of $Fe_3O_4$, leaving Au nanoparticle monolayer structure with gaps. Adapted with permission.[16] Copyright 2017, American Association for the Advancement of Science. f) Schematic of DNA-mediated gold cube monolayer on a template. g) SEM image of a gold cube monolayer array. Adapted with permission.[12b] Copyright 2015, American Chemical Society. h) Schematic of assembling Au nanoparticles into AAO nanopores driving by ultrasonication and corresponding SEM image. Adapted with permission.[44] Copyright 2019, Wiley-VCH.

### 3.2. Alignment of One-dimensional (1D) Nanomaterials

1D nanomaterials feature unique extraordinary electrical, optical, and thermal properties,[5, 63] and leveraging self-assembly techniques for preparing large-scale, monolayer, aligned 1D nanomaterials is crucial to bringing these properties at the nanoscale to macroscopic applications. 1D nanomaterials include nanowires and nanorods without quantum-confinement-induced physical properties, and nanotubes with quantum-confinement-induced physical properties for quantum-related applications. The alignment methods can be classified into two major categories: (1) force directed alignment methods, and (2) confinement-based alignment methods.

Force-directed alignment methods have been widely used to align a large variety of 1D nanomaterials. For example, Kim et al. developed a CdSe/CdS core/shell nanorod monolayer through air/liquid interface assembly.[64] The ligand density on nanorods were modified to manipulate the depletion force. The depletion attraction between nanorod and interface is stronger than that between two nanorods in a wide range of ligand density, leading to initial pinning of NRs at the interface. The subsequent alignment is determined by the relative magnitude of the attraction for nanorod-to-interface to nanorod-to-nanorod. For a dense ligand layer, the nanorod-to-interface attraction is slightly stronger than nanorod-to-nanorod, which causes the growth of a close-packed structure along both lateral and vertical directions from the nanorod absorbed at the interface (Figure 7a and b). Also, external force can direct the alignment. Kang et al. used stretching force to assemble the DNA nanowires with a parallel structure on PDMS substrate (Figure 7c).[65] As the receding meniscus moved due to the motion of PET plate, the anchored DNA was stretched and acted as the nucleation site. More individual DNA molecules were then allowed to continuously transport and accumulate into aligned nanowires on the flat PDMS substrate (Figure 7d). The fluorescence and AFM images show that resulted aligned DNA monolayer arrays feature excellent uniformity (Figure 7e). Similarly, during the spray-coating process, the spraying angle (i.e., gazing incidence) can be adjusted to guide the alignment of assembled nanomaterials.[45]

One most widely recognized nanotube is carbon nanotube (CNT), which has unique electronic and photonic properties for a variety of electronic and photonic applications;[66] see Section 4 for more details. CNTs can also been aligned through force-directed alignment techniques. For example, Joo et al. demonstrated a dose-controlled, floating evaporative self-assembly method to align the polymer-encapsulated semiconducting CNTs.[67] In this method, the hydrophobic substrate was immersed in the water and the semiconducting CNT suspension was dropped into the water in the vicinity of the substrate. Because of the surface tension difference, the CNT suspension droplets spread on top of water across the substrate surface. When the substrate was slowly withdrawn out of the liquid, aligned nanotubes were assembled on the substrate surface, indicating a self-assembly at the air-liquid-solid interface. The packing density and standard angle deviation with respect to the principal alignment direction of obtained aligned CNT films is ~50 CNTs/$\mu$m and ±14°, respectively. Furthermore, inspired by the "coffee ring effect" which describes that the particles in a droplet tend to gather at the edge of the droplet and form a ring-shape pattern after the liquid evaporates. T. Shastry et al. proposed a fast evaporation-driven self-assembly (EDSA) method for CNTs alignment by reducing the atmosphere pressure.[53] The evaporation process generated a friction force, transferring the particles toward the substrate-liquid contact line (i.e., assembly at the liquid-solid interface) and forming a densely packed assembly. The standard angle deviation of obtained aligned CNT films is ±5°. Moreover, K. Jinkins et al. improved the quality of obtained films by utilizing a 2D nematic tangential flow interfacial self-assembly (TaFISA), which confined and assembled CNTs at the interface between the CNT suspension and water (Figure 7f), indicating a liquid-liquid interface assembly.[68] As Figure 7g shows, by slowly lifting the substrate up from the liquid, CNTs flowing across it were assembled onto the substrate along the suspension flow direction. The packing density is close to 50 CNTs/$\mu$m and the standard angle deviation is ±6°. Moreover, Liu et al. self-assembled a suspension of semiconducting CNTs with ultra-high purity (>99.9999%) at the liquid-liquid interface (Figure 7h).[39] Instead of utilizing liquid flow, 2-butene-1,4-diol was used as top layer liquid since it has a strong affinity to the $SiO_2$ substrate. When the substrate was gently pulled out of the liquid, CNTs were confined at and aligned along the contact line at the interface of the wafer and the top layer liquid, and then assembled onto the substrate. This alignment technique was named dimension-limited self-alignment (DLSA) because it limited the rotation degrees of freedom of CNTs during the assembly process. Impressively, the DLSA method can produce high-purity, well-aligned, and high-density CNTs arrays on 4-inch wafer. The packing density is ~200 CNTs/$\mu$m and the standard angle deviation is ~±8°.

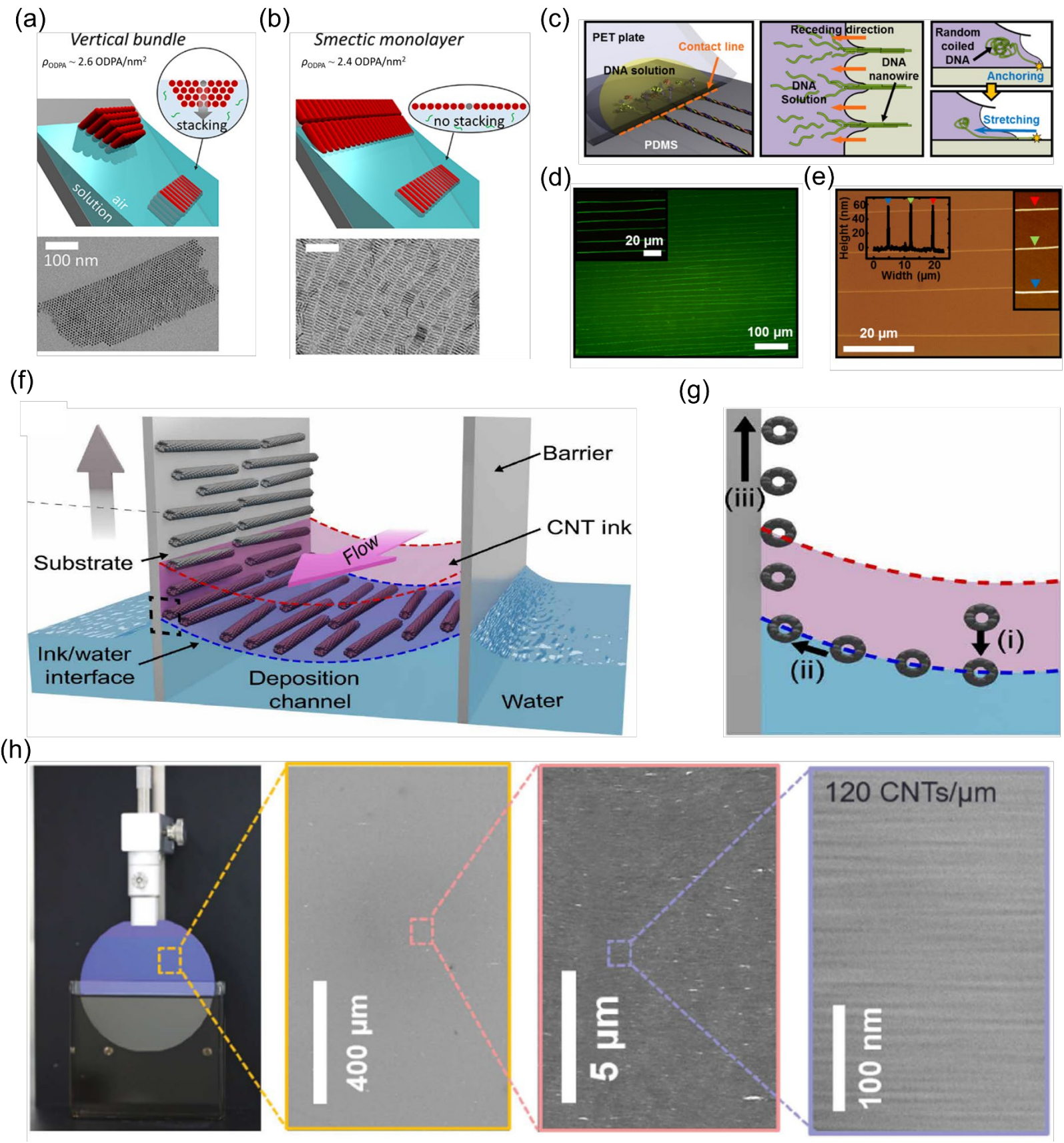

**Figure 7.** Force directed alignment of monolayer nanomaterials. Schematic of monolayer assembly of a) vertical and b) smectic nanorods with different octadecylphosphonic acid (ODPA) amount and corresponding TEM image. Adapted with permission.[64] Copyright 2019, American Chemical Society. c) Schematic illustration of the formation of highly aligned DNA nanowires on PDMS substrate by flow-assisted self-assembly. d) A representative fluorescent image of highly aligned DNA nanowires (green emitting) on the PDMS substrate, obtained at a moving speed of PET plate of 6 mm/min. e) AFM image of DNA array on the PDMS substrate. Reproduced with permission.[65] Copyright 2015, American Chemical Society. f) Illustration of TaFISA experimental setup and g) the side view of CNT suspension-water interface and self-assembly process. Adapted with permission.[68] Copyright 2021, American Association for the Advancement of Science. h) Optical image of the dip-coating setup for coating CNTs on a 4-inch silicon wafer and SEM images showing DLSA-prepared CNT array at different magnifications. Reproduced with permission.[39] Copyright 2020, American Association for the Advancement of Science.

Confinement-based alignment methods can also provide efficient alignment of 1D nanomaterials. Such confinement can be formed chemically. For example, phospholipid striped phases, which expose polar heads and nonpolar tails in alternating stripes, are designed on highly oriented pyrolytic graphite to trap the Au nanowires (Figure 8a-c).[5] The stripes of polar headgroups absorbed water to form nanometer-wide water channels, which allowed the Au nanowires to assemble in a well-aligned manner (Figure 8d-e). The confinement can also be formed through physical geometry, which needs the matchable shape and size of substrate traps and nanomaterials. For example, Flauraud et al. created a lithography-based trap pattern to assemble Au nanorods using an evaporation strategy (Figure 8f).[47] Solvent evaporation from the receding solvent/air interface, which can be accelerated by substrate heating, induces a convective flow that drags nanoparticles from the bulk to the surface of the suspension. The flow produces a dense accumulation of nanoparticles in the wedge-shaped region adjacent to the receding contact line and sustains the accumulation against nanoparticle back diffusion. Eventually, the matchable Au nanorods fit the traps and form an aligned monolayer array (Figure 8g). For CNTs, Cao et al. prepared aligned semiconducting CNTs with 99% purity at the interface of water and 1,2-dichloroethane (DCE) surface by creating the confinement through two moving bars. Specifically, as Figure 8h shows, when two mobile bars moved toward each other at the air/water interface, a uniaxial compress force was applied to align all CNTs along the direction perpendicular to the bar movement direction. The assembled CNTs were horizontally transferred onto the substrate through the Langmuir-Schaefer deposition. The obtained packing density is ~1,100 CNTs/$\mu$m and the standard angle deviation is ±17°.

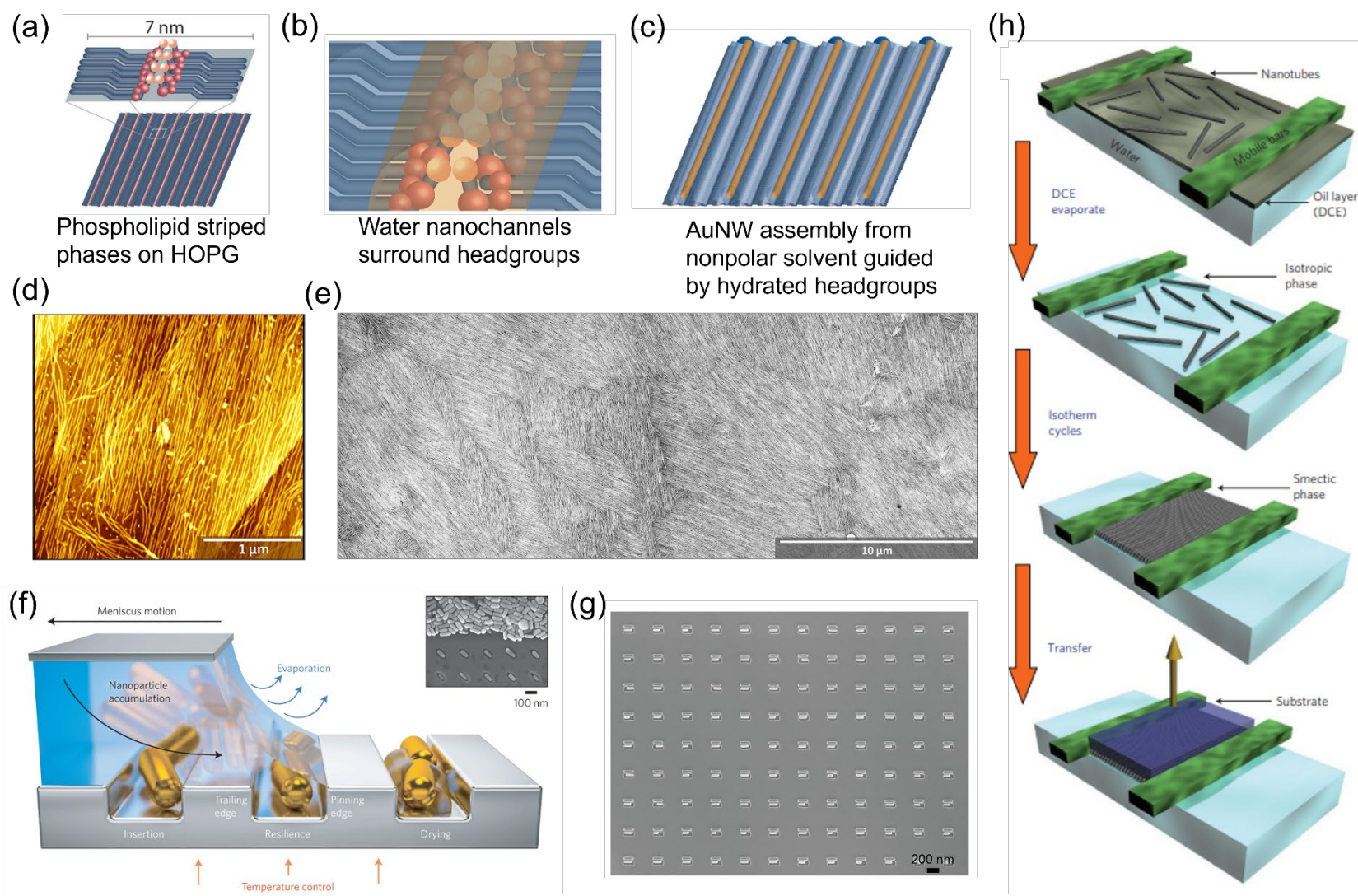

**Figure 8.** Confinement-based alignment of nanomaterials. a-c) Schematic of Au nanowires assembled in nanometer wide water channels. d) AFM images of Au nanowires deposited on diyne PE templates for 15 min. e) Large-area SEM image of aligned Au nanowires. Adapted with permission.[5] Copyright 2019, Elsevier. f) Schematic of the capillary assembly of nanorods onto traps of a low-wetting substrate. g) SEM image of Au nanorods (110 nm x 40 nm x 40 nm) selectively assembled in an array of 70 nm x 170 nm-wide, 28 nm-deep straight edged traps with single auxiliary sidewalls. Adapted with permission.[47] Copyright 2017, Nature Publication Group. h) Schematic illustration of the CNT alignment induced by the confinement created by moving two bars. Reproduced with permission.[69] Copyright 2013, Nature Publication Group.

### 3.3. Summary

The formation and properties of nanomaterial monolayer structures, such as nanomaterial gap and arrangement, are of paramount importance due to their substantial impact on monolayer characteristics. These structures, encompassing both close-packed and gapped arrangements, are primarily formed through self-assembly processes driven by interparticle forces such as van der Waals, electrostatic interactions, and capillary forces. The close-packed monolayers, characterized by their uniform particle size and shape, confer enhanced electrical and mechanical properties, making them ideal for diverse applications like structural coloration. In contrast, gapped monolayers with controlled interparticle gaps exhibit unique light-matter and electron-matter interactions, crucial for applications in high-sensitivity detection. The alignment of nanomaterials in these monolayers, achievable through external forces, chemical confinement, or geometric constraints, plays a critical role in defining their macroscopic properties.

## 4. Applications
### 4.1. Electronic Applications

Despite its immense success in past decades, the scaling of silicon-based field-effect transistors (FETs) and their integration density predicted by Moore's law starts to approach physical limits, such as issues of large leakage currents and heat dissipation when scaling down transistor dimensions, and the sustainability of Moore's law becomes more and more challenging. However, emerging applications, such as artificial intelligence, cryptocurrency, and virtual reality, urgently call for novel energy-efficient transistors with more computing resources. Hence, alternative material platforms, especially those at the nanoscale with quantum-confinement-induced properties, are extensively being explored. Among them, CNTs, which are one-dimensional (1D) nanostructures, exhibit high charge mobility and exceptional thermal conductivity and become one of the most promising candidates for next-generation FETs.[66]

Macroscopic assemblies of CNTs are required for the large-scale manufacturing of CNT-FETs and their integrated circuits. The degree of CNT alignment and the purity of semiconducting CNTs are two crucial factors for the FET performance, such as the FET mobility and on/off ratios. As shown in Figure 9, in order to catch up with the performance of an individual CNT FET and preserve the properties of nanoscale electronic devices in their macroscopic counterparts, CNT assemblies need to be aligned in an array and enriched with semiconducting types. The strong scattering between nanotube junctions in random network assemblies reduces mobility.[70] Furthermore, the synthesis of CNTs generally leads to a mixture of both metallic and semiconducting types. If metallic CNTs are not sufficiently removed and have residues, their formed percolation path can significantly decrease the on/off ratio.[71] Moreover, monolayer films with appropriate spacing between CNTs are preferred to reduce the screening effect for large on/off ratios.[71] Hence, the ideal assemblies will be aligned CNT arrays with an intermediate alignment density, such as 100-200 CNTs/$\mu$m, and a high semiconducting purity.[39]

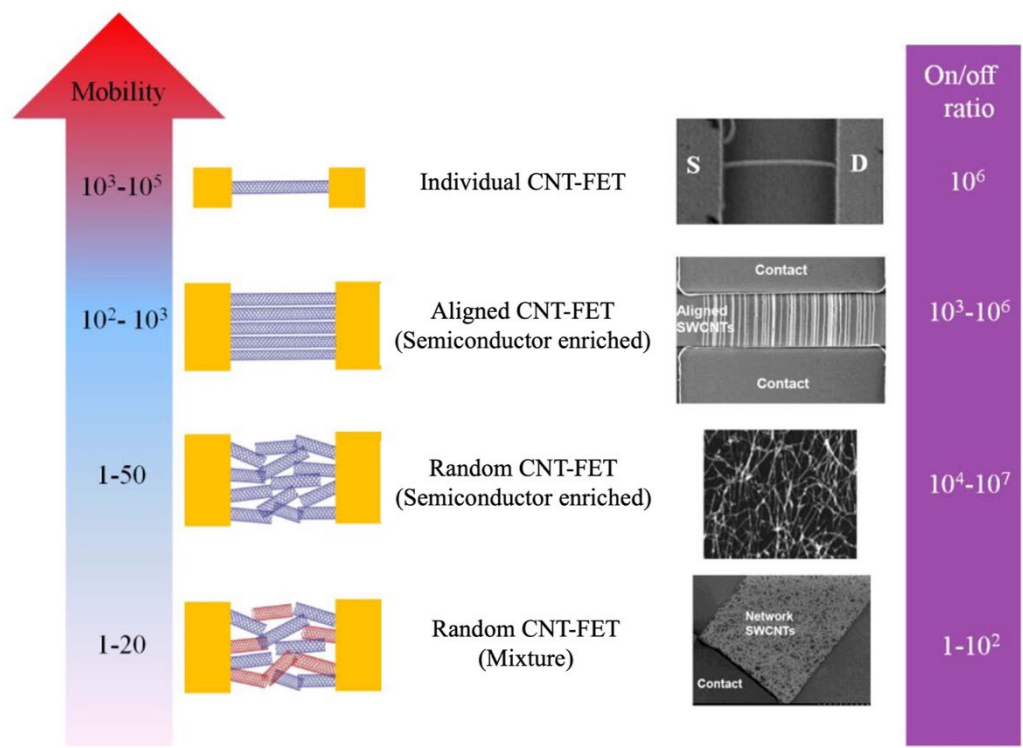

**Figure 9.** The mobility and on/off ratio of CNT-FETs made from CNT assemblies with different morphologies and purities of semiconducting types. Adapted with permission.[72] Copyright 2011, American Chemical Society.

 The post-growth self-assembly of CNTs is a promising technique to prepare such assemblies. Highly purified semiconducting CNT aqueous suspensions nowadays can be prepared from synthesized mixtures in a large variety of approaches,[73] such as DNA-assisted sorting, conjugated polymer-assisted sorting, density gradient ultracentrifugation, aqueous two-phase extraction, and gel chromatography.[74] These highly purified aqueous suspensions can be self-assembled into aligned monolayers, such as evaporation-driven self-assembly and the LB and LS self-assembly mentioned before. In this review paper, we will concisely describe a few recent works that demonstrate high-performance large-scale FETs based on self-assembled monolayers of aligned semiconducting CNTs. The detailed review of CNT-FETs is beyond the scope of this review and the readers can refer to other more comprehensive reviews, such as Cao's work.[75]

 By optimizing the concentration and flow rate of the CNTs suspension and the lifting speed of the substrate, K. Jinkins *et al.* fabricated highly aligned CNT films with intermediate density (∼50 CNTs/μm), meeting the requirement of CNT-FETs.[68] In order to study the performance of CNT-FETs with different alignment qualities, the authors also prepared aligned semiconducting CNT films using the floating evaporative self-assembly (FESA) method and fabricated FETs with both types of films (Figure 10a). Compared to the high-quality films prepared using the TaFISA method, the FESA method sometimes results in randomly oriented or overlapped CNTs. As shown in Figure 10b, the low alignment uniformity led to nonuniform electrical characteristics because disordered CNTs suppressed the on-current in CNT-FETs. Moreover, as mentioned before, the presence of metallic CNTs reduces the on/off ratios of FET, causes current leakage, and leads to stability and reliability issues over time. Hence, high-purity semiconducting CNT films are crucial for improving the performance of CNT-FETs. Liu et al. improved the purity of semiconducting CNT suspensions to over 99.9999% and fabricated aligned, high-density CNTs films for FETs by the DLSA method.[39] As shown in Figure 10c-d, the top-gate FETs fabricated from obtained films exhibited an on-state current over 1.3 mA/μm when setting the bias voltage of -1 V and an on/off ratio greater than $10^5$. Remarkably, as Figure 10e shows, the peak transconductance $g_m$ was 9 μS per CNT, which was the highest among all reported CNT film transistors. In addition, the authors fabricated ring oscillators (RO) utilizing CNT-FETs. The highest oscillating frequency reached 8.06 GHz and the corresponding stage delay was 12.4 ps. The power spectrum from this RO is shown in Figure 10f. Figure 10g compares the stage delay of this work with other works and the DLSA-prepared CNT-FET is the fastest among a variety of reported nanomaterial-based RO within similar electrical design and test conditions. Notably, based on this work, Lin et al. engineered the CNT-electrode contact type to achieve the low contact resistance of 90 Ω μm.[76] This innovation enables the creation of CNT-FETs with carrier mobility higher than silicon-based 10 nm metal-oxide-semiconductor FET.

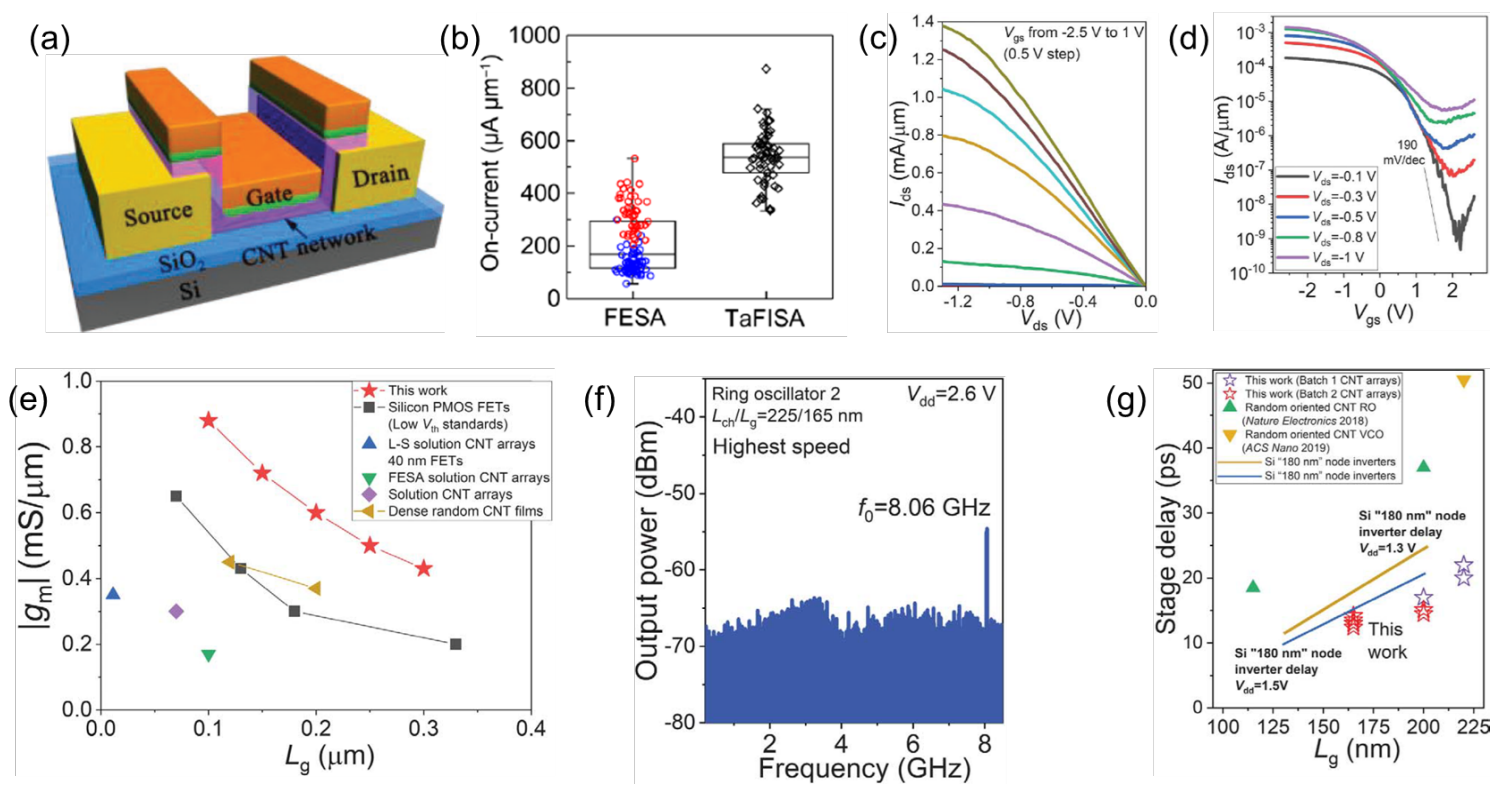

**Figure 10**. CNT-based FETs and the corresponding performance. a) Schematic diagram of a CNT-based top-gate FET. Reproduced with permission.[77] Copyright 2017, AIP Publishing. b) On-current comparison between FESA and TaFISA prepared FETs. Reproduced with permission.[68] Copyright 2021, American Association for the Advancement of Science. c) and d) The output characteristics and transfer characteristic of the CNT FET. e) Benchmark of transconductance ($g_m$) versus gate length ($L_g$) of DLSA-prepared CNT FET and other reported CNT FET and commercial Si PMOS transistor.[78]

f) Power spectrum from the RO with the highest oscillating frequency of 8.06 GHz. g) Benchmark of stage delay time versus $L_g$ of DLSA-prepared CNT FET RO and other CNT ROs with similar gate length.[78a, 78d, 79] Reproduced with permission.[39] Copyright 2020, American Association for the Advancement of Science.

In addition to electronic applications, the extraordinary optical properties of CNTs lead to promising photonic applications as well. Specifically, semiconducting CNTs possess broadband optical responses (0.2-1.5 eV) and exhibit ultra-high electronic carrier mobility and they become one of the ideal material platforms for CMOS-compatible optoelectronic integrated circuits (OEICs), which seamlessly merge electronic and optical functionalities on a single chip. Y. Liu et al. proposed a monolithic 3D OEIC system consisting of a photovoltaic receiver and an electrical-driven transmitter, which are both fabricated using thin films of CNTs prepared through the liquid-phase deposition method.[80] The photoreceiver comprises a photodetector and a signal processor. Typically, these two components are manufactured using different materials, such as gallium arsenide for the photodetector and silicon for the electrical circuits. In contrast, as Figure 11a shows, these two components can be implemented using semiconducting CNT films and integrated. When the cascaded photodetector is illuminated with infrared light, it generates a photovoltage that controls the subsequent integrated n-type FET, thereby realizing the optical gate function (Figure 11b). Notably, the CNT-based photodetector offers a broadband window covering the entire near-infrared band (1165- 2100 nm) and both the electrical and optical on/off ratio can both reach $10^5$. The structure of the transmitter resembles that of the photodetector and both consist of one transistor and one electro-optical device. The difference is that the photoreceiver consists of a photodetector connected to the gate of the transistor while the transmitter consists of a photoemitter connected to the drain of the transistor (Figure 11c). The electroluminescence spectrum is narrow by using high-purity semiconducting CNT films and the emission band can also be controlled by choosing CNT chiralities. As shown in Figure 11d, the light emission from a CNT-based transmitter can be switched on and off under a modulated electrical control signal.

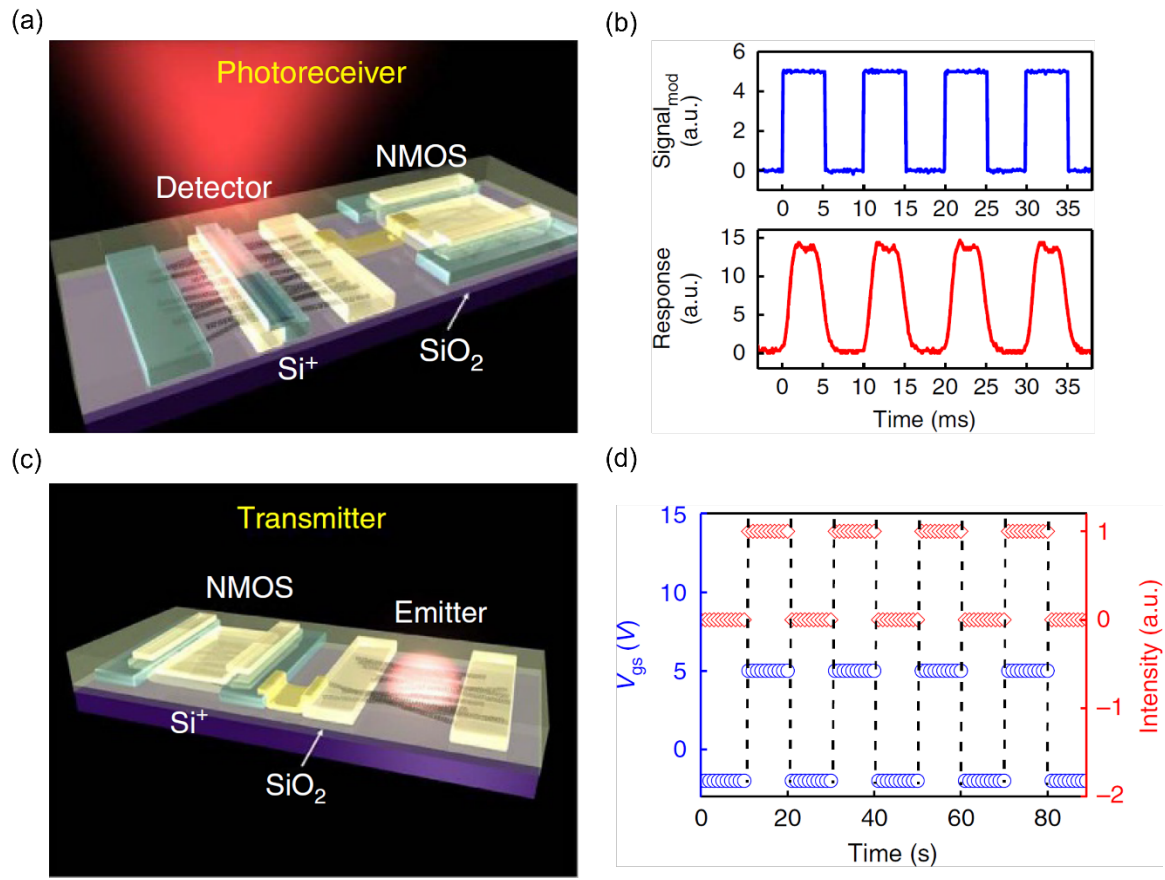

**Figure 11.** a) Schematic of CNT-based photoreceiver. b) Dynamic time trace of the photoreceiver. c) Schematic of CNT-based transmitter. d) Time trace of the transmitter. Reproduced with permission.[80] Copyright 2017, Nature Publication Group.

## 4.2. Photonic Applications

*4.2.1. SERS Sensor*

SERS is a powerful analytical technique that enhances the Raman scattering effect of molecules absorbed on or near certain noble metal surfaces or nanomaterials, typically gold, silver, and platinum.[81] This enhancement enables the detection of trace concentrations of pollutant molecules. The phenomenon behind SERS is primarily attributed to two main mechanisms: electromagnetic enhancement and chemical enhancement. Electromagnetic enhancement occurs when the incident light induces localized surface plasmon resonances in the metal substrate, resulting in a significant increase in the electric field near the surface.[82] This amplified electric field boosts the Raman scattering of the molecules in close proximity. Chemical enhancement, though less dominant, involves charge transfer interactions between the molecule and the metal surface, which can also increase the Raman signal.[83]

Monolayer nanomaterials as SERS substrates exhibit significant advantages, including extraordinary sensitivity and selectivity, uniform hotspot distribution, tunable structures and shapes, along with excellent chemical stability. These nanomaterials can be surface functionalized to enhance affinity towards specific analytes, thereby increasing selectivity. Their diminutive size and flexibility facilitate easy integration into wearable sensors and microfluidic chips, enabling real-time monitoring and on-site detection. For example, Zhu et al. fabricated a wearable SERS sensor based on PS microparticle monolayer with Ag nanoparticle monolayer decoration (Figure 12a).[84] This hierarchical structure provides huge hotspots between Ag nanoparticles (nanoparticle gaps: 5 nm), which enhances the electromagnetic fields for sensitive molecule detection (Figure 12b). Correspondingly, the authors detected R6G molecules using the monolayer SERS substrate, exhibiting a good linear response of R6G from $10^{-5}$ M down to $10^{-13}$ M (Figure 12c-d). It is also well known that sub-1-nm nanogaps between two metallic nanomaterials demonstrate highly intense and localized electromagnetic fields, resulting in a large Raman enhancement for single-molecule detection.[85] For example, Si et al. developed a monolayer gold nanoparticles through liquid-liquid interface assembly.[86] By manipulating the diameters of Au nanoparticles (30 nm and 120 nm), the average nanoparticle gaps can be controlled to 1.37 nm and 0.48 nm (Figure 12e and g). As the results of the Raman signal response over R6G molecule, the sub-1-nm gaps behave much higher Raman intensity of R6G molecules (Figure 12f and h). As the nanoparticle gaps are super small, the interactions among nanoparticle monolayer is strong enough to support mechanical cycling without sacrificing its SERS performance (Figure 12i). The simulation also indicates the amplification of electromagnetic fields between Au nanoparticles (Figure 12j). Quantum mechanical effects emerge on plasmonic structures with such small geometry, especially for sub-1-nm gaps that are parallel to the electric field polarizations of the incident light, because free electrons are able to tunnel through the gap, which would effectively increase the local field enhancement.[87]

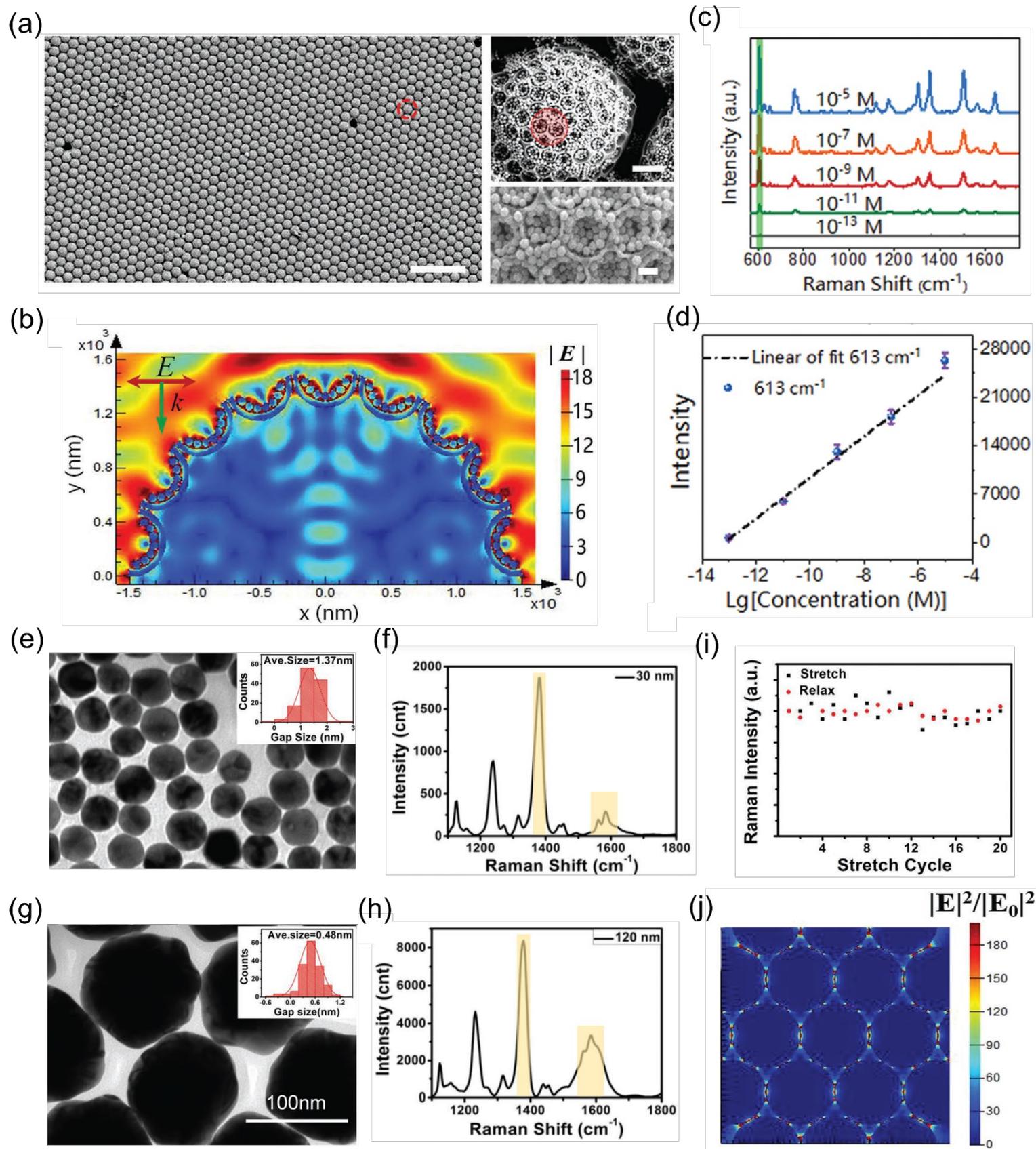

**Figure 12.** a) SEM image of Au@SiO$_2$ monolayer nanoparticles deposited by Ag nanoparticles (Au@SiO$_2$-Ag) and enlarged circled area. Scale bars, 20 μm (left), 1 μm (right top), 200 nm (right bottom). b) Simulated electromagnetic field of Au@SiO$_2$-Ag monolayer structure excited at 633 nm. c) SERS spectra of R6G molecular absorbed on Au@SiO$_2$-Ag monolayer with different concentrations excited at 633 nm and d) summarized concentration-dependent SERS intensity at 613 cm$^{-1}$. Reproduced with permission.[84] Copyright 2022, Wiley-VCH. e) TEM image of Au nanoparticle (30 nm in diameter) monolayer and average interparticle gap (inset), and corresponding f) SERS spectrum of rhodamine 6G (R6G) absorption. g) TEM image of Au nanoparticle (120 nm in diameter) monolayer and average interparticle gap (inset), and corresponding h) SERS spectrum of rhodamine 6G (R6G) absorption. Note that the R6G concentration is 1×10$^{-10}$ M, the laser wavelength is 633 nm, and the shadowed peaks are related to R6G molecular. i) SERS performance of monolayer on PDMS after 20 stretch/relax cycles. j) Simulated electrical field intensity distribution for 120 nm Au nanoparticle excited by 633 nm wavelength laser. Reproduced with permission.[86] Copyright 2016, Wiley-VCH.

*4.2.2. Multi-stack Engineering for Photonic Applications*

Monolayer films are generally too thin for sufficient light-matter interaction. Hence, creating multilayer stacks in a controlled and engineering manner is crucial for many photonic applications. Here, we take CNTs as an example to describe techniques for preparing multilayers and stacks. Within the post-growth solution-based CNT self-assembly techniques, solution-based controlled vacuum filtration technique invented by He et al. is a preferred approach to highly-aligned, densely-packed and chirality-enriched CNT films with controllable thickness for photonic applications.[88] Figure 13a displays the vacuum filtration system for assembling individual CNTs in well-dispersed aqueous suspensions. As water molecules pass through the filter membrane at a controlled low filtration speed, the concentration of CNT suspensions on top of the membrane increases and CNTs interact and align with each other because of liquid crystal phase transition. By controlling the volume of filtrated suspension, the obtained film thickness can be adjusted from a few nanometers to a few hundred nanometers. Figure 13a displays the photo and scanning electron microscopy images of prepared CNT films. The volume packing density is ~3.8×10$^5$ tubes in a cross-sectional area of 1 μm$^2$) and the standard angle deviation ~±1.5°. Furthermore, this technique is universally applicable to any CNT suspensions, particularly chirality-enriched CNT suspensions. Hence, the controlled vacuum filtration method can produce aligned CNT films with arbitrary CNT chirality, enabling a variety of photonic applications.

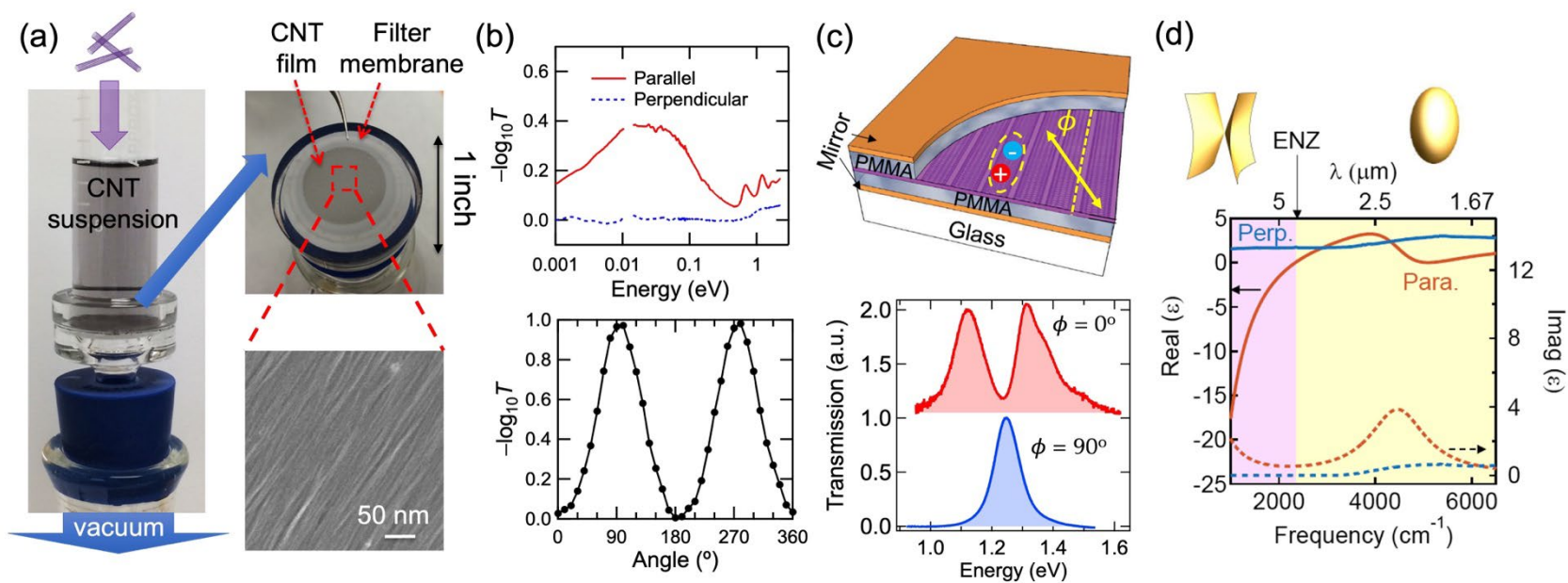

**Figure 13.** a) The CNT suspension goes through the vacuum filtration system and the wafer-scale aligned films formed in this process. b) Broadband attenuation spectra for the light polarization parallel and perpendicular to CNT alignment direction (Top panel) and angle-dependent attenuation in the terahertz range (Bottom panel). Adapted with permission.[88] Copyright 2016, Nature Publication Group. c) Schematic of an aligned semiconducting CNT film that is embedded in a microcavity (Top panel), which displays angle-dependent anisotropic exciton-polaritons. Reproduced with permission.[89] Copyright 2018, Nature Publication Group. d) Experimentally measured dielectric functions in parallel and perpendicular to the CNT alignment direction. Reproduced with permission.[90] Copyright 2019, American Chemical Society.

The tubular structure of CNTs leads to the formation of subbands and sharps electronic densities of states called van Hove singularities. The 1D nature of CNTs features strongly anisotropic optical absorption. When the light polarization is parallel to the CNT alignment direction, strong excitonic transitions between subbands occur. In contrast, for perpendicular polarization, optical transitions with different optical selection rules can be excited but strongly suppressed because of the depolarization effect. All of these anisotropic optical properties in individual CNTs are preserved in wafer-scale aligned films prepared using the controlled vacuum filtration technique. The top panel in Figure 13b shows the broadband attenuation spectra of such a film and the bottom panel displays the polarization angle-dependent attenuation in the terahertz range. Furthermore, Gao et al. demonstrated a unique microcavity exciton-polariton architecture, which can operate in weak or strong coupling regions depending on the light polarization, by embedding aligned semiconducting single-chiral (6,5) CNT film into a Fabry-Pérot cavity (Figure 13c top panel).[89] When changing the angle between probe light polarization and alignment direction, the coupling strength between zero (i.e., weak coupling) and maximum value (i.e., strong coupling) can be selected on demand. The dispersion of the exciton-polariton featured the existence of exceptional points. As the aligned film thickness increases, the coupling strength showed a cooperative enhancement. Moreover, Gao et al. demonstrated that aligned CNT films with a mixture of semiconductors and metals not only possessed strongly anisotropic dielectric functions in the mid-infrared range at high temperatures but also featured a topological transition from an open-geometry isofrequency contour to a closed-geometry isofrequency contour when the real part of the dielectric constant along CNT alignment direction crossed zero (Figure 13d).[90] The zero-crossing frequency is called epsilon-near-zero (ENZ) point. Based on such a property, a spectrally selective thermal emitter based on aligned CNT films and subwavelength-sized cavities was demonstrated. Especially, the smallest cavity with a volume about $\lambda^3/700$ was successfully obtained and the enhancement of photon density of states compared to blackbody radiation was >100.

In addition, multiple aligned CNT films can be stacked and engineered to further enhance the anisotropic optical response or bring new functionality for controlling light-matter interaction. N. Komatsu et al. proposed a doping-engineered 3D aligned CNT architecture through layer stacking and chemical doping.[91] As shown in Figure 14a, a 2-inch aligned CNT film was cut into pieces and one piece was transferred onto a substrate. The transferred film was then immersed into a nitric acid bath to be made strongly *p*-doped. This transfer-doping process was repeated multiple times. The extinction ratio for linearly polarized attenuation in the terahertz range linearly increased with respect to the increasing number of stacked layers (Figure 14b). Instead of keeping the alignment direction the same when stacking multiple layers, J. Doumani et al. demonstrated that when stacking aligned CNT films with twist angles, the twist-stacked multilayer film displayed circular dichroism (CD).[92] As shown in Figure 14c, for a single-layer sample, the CD is negligible. In contrast, when stacking second and third layers with a fixed rotation angle of 30º, the CD was gradually enhanced. Furthermore, the sign of CD was controlled by the rotation direction. This structural-induced chiro-optical effect in the twist-stacks of aligned CNTs is promising for applications in biosensing, secure communication, quantum information processing, and spintronics.

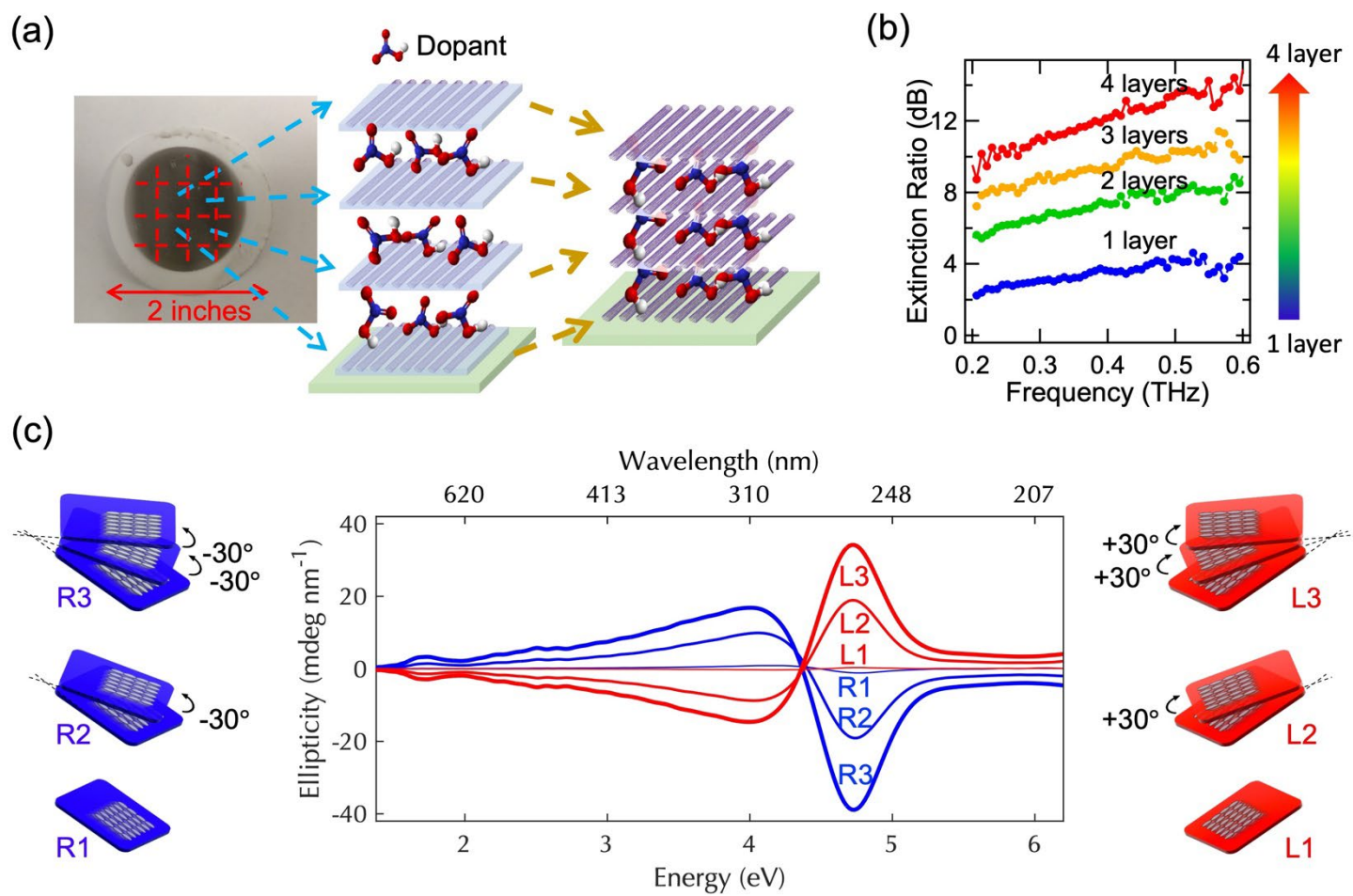

**Figure 14.** a) Schematic of the aligned CNT multilayers fabricated via stacking and doping method. b) Diagram of extinction ratio versus frequency for different stacked CNT layers. Reproduced with permission.[91] Copyright 2017, Wiley-VCH. c) Twisted stacks of aligned CNT films with different twist directions and corresponding circular dichroism spectra. Reproduced with permission.[92] Copyright 2023, Nature Publication Group.

### 4.3. Others

*4.3.1. Memory Device*

In the contemporary landscape of electronic systems, memory devices are indispensable for data storage and retrieval, playing an important role in data processing and storage.[93] The integration of monolayer nanomaterials into memory devices presents a plethora of advantages.[93-94] The integration of monolayer nanomaterials into memory devices presents a plethora of advantages.[94] The minuscule dimensions and extensive surface area of these nanomaterials markedly boost storage density, allowing for augmented data storage within a compact footprint. Furthermore, the unique physical and chemical properties of monolayer nanomaterials, such as enhanced electrical conductivity and robust charge storage capabilities, provide a solid foundation for swift and efficient read-write cycles.

Additionally, the versatility of these materials permits customized design and optimization tailored to specific memory device applications. For example, Wang et al. who engineered an organic nano-floating-gate memory (NFGM) device using a AuNPs@PS superlattice monolayer (Figure 15a).[14d] This innovative device showcased typical p-type field-effect behavior in its initial transfer and output characteristics (Figure 15b). It exhibited exceptional performance metrics, with a mobility of 0.72 cm$^2$ V$^{-1}$ s$^{-1}$ and an ON/OFF current ratio of 6.9x10$^5$. These figures are indicative of a substantial memory window, as well as high-speed programming and erasure capabilities. Moreover, the device's threshold voltage ($V_{TH}$) shifts were scrutinized to assess its capability for hole-electron trapping, revealing extensive $V_{TH}$ shifts in both hole trapping (20.1 V) and electron trapping modes (53.5 V) (Figure 15c-d). These results underscore that the ordered periodic structures of the superlattice monolayer significantly amplify its hole injection and electron trapping efficacies, thus illustrating the profound impact of monolayer nanomaterials in enhancing the functionality and efficiency of memory devices.

*4.3.2 Osmotic Energy Harvesting*

Osmotic energy harvesting represents a cutting-edge domain in renewable energy technologies, capitalizing on the osmotic pressure differential between freshwater and saltwater to generate electricity.[95] The innovative approach leverages a semi-permeable membrane, which permits water molecules to migrate from a region of lower salt concentration (freshwater) to higher concentration (brine). This migration creates water pressure, propelling the liquid through the membrane under pressure, and thereby inducing liquid flow.

Monolayer nanomaterial structures, characterized by their extensive surface area, adjustable channel size, and ultra-thin thickness, offer a suite of advantages. These include heightened surface activity, augmented ion selectivity, and rapid ion transport. These attributes render monolayer nanomaterials as an exemplary candidate for osmotic energy harvesting, facilitating high-power energy generation. For example, Xiao et al. adopted an air/liquid interfacial method to fabricate a monolayer of metal-organic framework (MOF) nanoparticles, subsequently integrating this layer onto an AAO substrate (Figure 15e).[96] This MOF monolayer exhibits enhanced ionic conductivity in the presence of increased ion concentration in the solution (Figure 15f-g), a property that significantly bolsters the efficiency of osmotic energy harvesting. Furthermore, when the osmotic energy harvesting device (Figure 15h) was tested under a salinity gradient, the MOF monolayer nanoparticles on the AAO substrate demonstrated a substantially higher power density compared to bare AAO substrates (Figure 15i). These findings underscore the potential of MOF monolayer nanoparticles in revolutionizing the field of osmotic energy harvesting, paving the way for more efficient and powerful energy generation technologies.

*4.3.3. Lithium-ion Battery*

The landscape of anode materials for lithium-ion batteries (LIBs) is evolving rapidly, driven by the quest for extended longevity and heightened energy densities. Nonetheless, the pursuit of prolonged cycling and augmented lithium storage capacities is frequently marred by substantial volumetric alterations. These changes pose a significant challenge, as they can lead to the pulverization of electrode materials during the processes of lithiation and delithiation.[97]

From a technological perspective, the inherent two-dimensional (2D) close-packed configuration of nanoparticle superlattices emerges as a pivotal innovation. This arrangement not only facilitates electron transport but also efficaciously mitigates the excessive side reactions between electrolytes and nanoparticles. Such attributes are critical for attaining superior electrochemical performances.[98] A notable advancement in this domain has been made by Li et al.,[99] who ingeniously crafted a $Fe_3O_4$ nanoparticle assembly into a monolayer supertube (Figure 15j). This structure is particularly well-suited for LIB applications. The close-packed nanoparticle arrangement exhibits remarkable electrochemical stability. In contrast, for the gapped nanoparticle monolayer (Figure 15k), the nanoparticles deviate from their original cubic geometry after 50 cycles of lithiation and delithiation at a current of 0.2 A $g^{-1}$. Impressively, the monolayer nanoparticle structure showcases significantly higher average capacities compared to its super-rod counterparts across various current rates (Figure 15l). This exceptional rate performance is primarily attributed to the monolayer geometry, which substantially enhances lithiation and delithiation kinetics due to the reduced Li-ion diffusion pathways. Most strikingly, the $Fe_3O_4$ monolayer demonstrates an ability to maintain a stable capacity of 800 mAh $g^{-1}$, even after 500 cycles at an elevated current of 5 A $g^{-1}$ (Figure 15m). The overall electrochemical performance of the $Fe_3O_4$ monolayer structure outshines most previously reported $Fe_3O_4$-based anode materials,[99] marking a significant stride in the development of advanced anode materials for lithium-ion batteries.

*4.3.4. Polymer Solar Cell*

In the realm of solar cells, the deployment of monolayer nanoparticles is pivotal, particularly in augmenting the efficacy of polymer solar cells (PSCs). The distinction between monolayer structures and their multilayer counterparts lies in several advantages. Monolayers offer a more homogeneous and denser interface, which is instrumental in elevating charge transfer efficiency. This is primarily achieved by mitigating the recombination of electron-hole pairs, a critical aspect in the amelioration of overall solar cell performance.[100] The homogeneity inherent in monolayer structures is exceedingly advantageous for the effective segregation and translocation of charges. Quantum dot monolayers, serving as potent sensitizers, encompass an extensive spectral range extending from visible light to the near-infrared spectrum. This capability significantly enhances the light absorption proficiency of solar cells. Furthermore, these monolayers are adept at generating multiple excitons from a singular photon via mechanisms such as carrier multiplication (CM) or multiple exciton generation (MEG), thereby enhancing the photovoltaic conversion efficiency.[101]

A typical example of this is the work by Moon et al.,[102] who developed a high-performance polymer solar cell integrated with a CdSe quantum dot monolayer (Figure 15n-p). Empirical data from the absorption spectra of this PSC device, equipped with a CdSe quantum dot monolayer, exhibit an approximate 10% increase in light absorption at a 550 nm wavelength, in contrast to other structures (Figure 15q). Simulations further elucidate that the CdSe quantum dot monolayer manifests a surface plasmon resonance (SPR) effect (Figure 15r). This phenomenon engenders robust electromagnetic fields, which in turn amplify the light absorption capabilities of the PSC device. Notably, with the incorporation of the CdSe quantum dot monolayer, the internal quantum efficiency (IQE) spectrum of the PSC device is significantly enhanced (Figure 15s). This indicates that the absorbed photons lead to the generation of separated charge carrier pairs, with a majority of the photogenerated carriers being efficaciously collected at the electrodes, thus circumventing recombination effects.

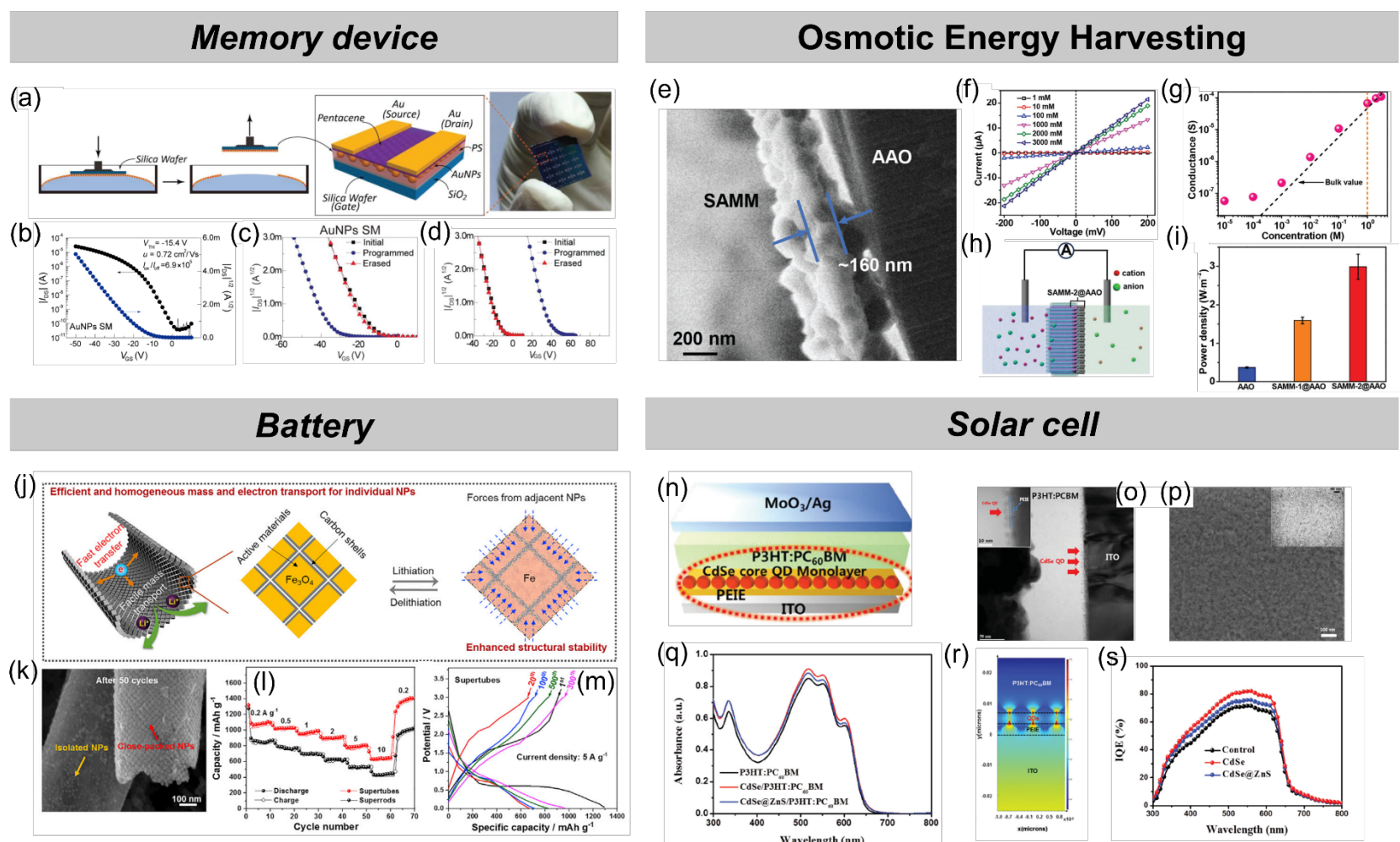

**Figure 15.** a) Schematic of gold nanoparticle monolayer on air-liquid interface transferred onto Si wafer and configuration of organic nano-floating-gate memory (NFGM) device. b) Representative transfer ($V_{DS}$=-30 V) and output characteristics of the pentacene-based organic NFGM devices with Au nanoparticle monolayer. c) Programming and erasing characteristics of the pentacene-based NFGM devices with Au nanoparticle monolayer in hole trapping mode

and d) electron trapping mode. Reproduced with permission.[14d] Copyright 2018, Wiley-VCH. e) Cross-section image of the MOF nanoparticle (i.e., SAMM-2) monolayer on anodic aluminum oxide (AAO) substrate. f) I−V curves of SAMM-2@AAO under NaCl electrolyte solutions with different concentrations. g) Transmembrane ionic conductance of SAMM-2@AAO with different salt concentrations. h) Osmotic energy-harvesting device under a salinity gradient. i) The maximum output power densities of AAO, SAMM-1@AAO, and SAMM-2@AAO harvested under a 500 mm/10 mm NaCl gradient. Reproduced with permission.[96] Copyright 2023, Wiley-VCH. j) Schematic of the lithiation/delithiation of $Fe_3O_4$ monolayer supertubes. k) SEM images of $Fe_3O_4$ monolayer after 50 cycles at 0.2 A g$^{-1}$. l) Rate performance of $Fe_3O_4$ monolayer supertubes and super-rods. m) Galvanostatic charge/discharge voltage profiles of $Fe_3O_4$ monolayer supertubes at different cycles. Reproduced with permission[99]. Copyright 2019, Elsevier. n) Inverted PSC structure with the CdSe quantum dot monolayer. o) Cross-sectional TEM image of the CdSe monolayer in the PSC structure. p) SEM image of CdSe quantum dot covered by the polyethylenimine ethoxylated (PEIE) polymer layer and a magnified image of the CdSe quantum dot/PEIE layer (inset). q) Absorption spectra of polymer films with and without a CdSe (or CdSe@ZnS) quantum dot monolayer. r) Simulated electric field profiles of PSC device with CdSe quantum dot monolayer at a wavelength of 550 nm. s) Internal quantum efficiency (IQE) spectra of PSC devices. Reproduced with permission.[102] Copyright 2015, Wiley-VCH

## 4. Conclusion and outlook

To achieve high performance and reliable devices of nanomaterials, the number of layer matters. MAN represents the highest controllability. This paper contributes a systematic summary of the solution-based assembly methods to achieve MAN, strategies to achieve highly ordered nanoparticle crystals and aligned MAN structures, and the applications of MAN in electronics and photonics. Despite the significant progress, it is important to point out the underlying challenges and future directions. The first challenge comes from the challenge of nanomaterial manufacturing. To achieve highly ordered MAN, ideally, monodispersed nanomaterials (identical shape and size) is required. At least, near-monodispersed nanoparticles with very narrow shape and size distribution are needed. Although some near-monodispersed nanoparticles are commercially available, the price is extremely high (> $100/g) and choices of nanomaterials are extremely limited (e.g., Au, and $SiO_2$). Therefore, scalable and cost-effective manufacturing methods for monodispersed nanomaterials are desired. Second, many MAN processes involve volatile organic solvents and developing eco-friendly MAN assembly methods will be an important direction. Third, the potential of MAN has not been fully explored which can be partly traced back to the limited choices of nanomaterials with narrow shape and size distribution. We hope this review will attract more attention to MAN manufacturing and devices and pave the way for the future development of MAN.

## Acknowledgment

L.Z. and B.L. were supported in part by the National Science Foundation (Grants No. 2003077, 2221102, and 2018852), PA Manufacturing Fellows Initiative, Sport & Performance Engineering Seed Grant of College of Engineering, Villanova University. J.F. and W.G. acknowledge the support from the National Science Foundation through Grants No. 2230727. W.G. acknowledges the support from the National Science Foundation through Grant No. 2321366. C.G. acknowledges the support from the Dean's Fellow of College of Engineering, Villanova University. A.D. acknowledges the support from ME Junior Research Scholar, Department of Mechanical Engineering, Villanova University. L. Zhao and J. Fan contributed equally to this work.

## Conflict of Interest

The authors declare no conflict of interest.

## Author Contributions

L. Z. performed conceptualization, validation, formal analysis, investigation, data curation, and wrote the original draft, and revised the final manuscript. J. F. performed validation, formal analysis, data curation, and wrote the original draft, and revised the final manuscript. C. G. performed validation, formal analysis, data curation. A. L. performed investigation and data curation. W. G. performed formal analysis, investigation, data curation, funding acquisition, resources, and wrote the original draft, and revised the final manuscript. B. L. performed formal analysis, investigation, data curation, funding acquisition, resources, and wrote the original draft, and revised the final manuscript.

## Authors' biography

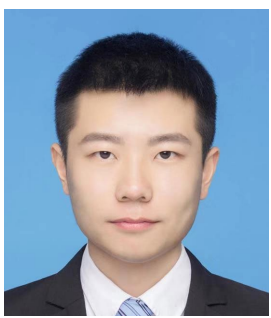

Liang Zhao is a Ph. D. candidate in the Department of Mechanical Engineering at Villanova University. He received his M.S. degree in the Department of Materials Science and Technology at Wuhan Textile University in 2019. His research interests focus on the assembly of various nanomaterials (1D-3D) and corresponding applications such as structural coloration, thermal management, and surface enhanced Raman spectroscopy.

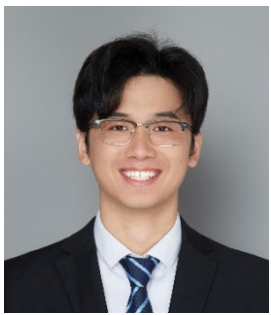

Jichao Fan is a Ph. D. candidate in the Department of Electrical and Computer Engineering at the University of Utah. He received his M.S. degree in physics from Stevens Institute of Technology in 2020. His research interests are in fundamentals of light-matter interaction of two-dimensional materials like circular dichroism, optical emission and electro-optic absorption.

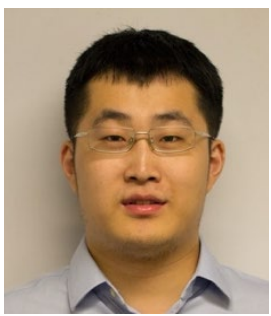

Dr. Weilu Gao is an Assistant Professor in the Department of Electrical and Computer Engineering at the University of Utah, United States. He received his B.S. degree in electrical engineering from Shanghai Jiao Tong University in 2011 and his Ph.D. degree in electrical and computer engineering from Rice University in 2016. His research interests are in photonics and optoelectronics of nanomaterials, including single-wall carbon nanotubes and two-dimensional materials, spanning fundamental research to applications in health, energy, imaging, sensing, computing, and communication.

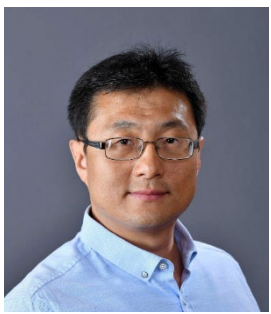

Dr. Bo Li is an Associate Professor in the Department of Mechanical Engineering at Villanova University, United States. He serves as the principal investigator of the Hybrid Nano-architecture and Advanced Manufacturing Lab (HNAM). Dr. Li obtained his both B.S. and M.S. degree in Polymer Engineering, from Sichuan University, China, and his Ph.D. degree in Mechanical Engineering from Northeastern University, United States, in 2013. His research interests include: 1) advanced manufacturing of nanomaterials and their hybrid architectures, 2) engineering micro-to-nano-scale interface between nanomaterials and substrates, 3) flexible and wearable electronics, and 4) multifunctional coatings and composites.

Reference

[1] a) P. J. Santos, P. A. Gabrys, L. Z. Zornberg, M. S. Lee, R. J. Macfarlane, *Nature* **2021**, *591*, 586-591; b) M. S. Lee, D. W. Yee, M. Ye, R. J. Macfarlane, *J. Am. Chem. Soc.* **2022**, *144*, 3330-3346; c) L. Xu, W. Ma, L. Wang, C. Xu, H. Kuang, N. A. Kotov, *Chem. Soc. Rev.* **2013**, *42*, 3114-3126.

[2] D. Liu, W. Cai, M. Marin, Y. Yin, Y. Li, *ChemNanoMat* **2019**, *5*, 1338-1360.

[3] G. Yang, J. Nanda, B. Wang, G. Chen, D. T. Hallinan Jr, *ACS Appl. Mater. Interfaces* **2017**, *9*, 13457-13470.

[4] J. Zhu, M. C. Hersam, *Adv. Mater.* **2017**, *29*, 1603895.

[5] A. G. Porter, T. Ouyang, T. R. Hayes, J. Biechele-Speziale, S. R. Russell, S. A. Claridge, *Chem* **2019**, *5*, 2264-2275.

[6] K. Matsuba, C. Wang, K. Saruwatari, Y. Uesusuki, K. Akatsuka, M. Osada, Y. Ebina, R. Ma, T. Sasaki, *Sci. Adv.* **2017**, *3*, e1700414.

[7] X. Li, L. Chen, Y. Ma, D. Weng, Z. Li, L. Song, X. Zhang, G. Yu, J. Wang, *Adv. Funct. Mater.* **2022**, *32*, 2205462.

[8] K. Volk, J. Fitzgerald, M. Retsch, M. Karg, *Adv. Mater.* **2015**, *27*, 7332-7337.

[9] H. Liu, M. Siron, M. Gao, D. Lu, Y. Bekenstein, D. Zhang, L. Dou, A. P. Alivisatos, P. Yang, *Nano Res.* **2020**, *13*, 1453-1458.

[10] aA. Toor, T. Feng, T. P. Russell, *Eur. Phys. J. E* **2016**, *39*, 57; bS. Hou, L. Bai, D. Lu, H. Duan, *Accounts Chem. Res.* **2023**, *56*, 740-751.

[11] H. Zhang, J. Cadusch, C. Kinnear, T. James, A. Roberts, P. Mulvaney, *ACS Nano* **2018**, *12*, 7529-7537.

[12] a) L. Zhao, B. Sidnawi, J. Fan, R. Chen, T. Scully, S. Dietrich, W. Gao, Q. Wu, B. Li, *ACS Appl. Mater. Interfaces* **2022**, *14*, 46095-46102; b) Q.-Y. Lin, Z. Li, K. A. Brown, M. N. O'Brien, M. B. Ross, Y. Zhou, S. Butun, P.-C. Chen, G. C. Schatz, V. P. Dravid, *Nano Lett.* **2015**, *15*, 4699-4703; c) Q. Y. Lin, J. A. Mason, Z. Li, W. Zhou, M. N. O'Brien, K. A. Brown, M. R. Jones, S. Butun, B. Lee, V. P. Dravid, *Science* **2018**, *359*, 669-672.

[13] a) L. Song, B. B. Xu, Q. Cheng, X. Wang, X. Luo, X. Chen, T. Chen, Y. Huang, *Sci. Adv.* **2021**, *7*, eabk2852; b) C. García Núñez, W. T. Navaraj, F. Liu, D. Shakthivel, R. Dahiya, *ACS Appl. Mater. Interfaces* **2018**, *10*, 3058-3068; c) T. Wen, S. A. Majetich, *ACS Nano* **2011**, *5*, 8868-8876.

[14] a) H. J. Seo, W. Jeong, S. Lee, G. D. Moon, *Nanoscale* **2018**, *10*, 5424-5430; b) X. D. Tian, Y. Lin, J. C. Dong, Y. J. Zhang, S. R. Wu, S. Y. Liu, Y. Zhang, J. F. Li, Z. Q. Tian, *Adv. Opt. Mater.* **2017**, *5*, 1700581; c) Q. Guo, M. Xu, Y. Yuan, R. Gu, J. Yao, *Langmuir* **2016**, *32*, 4530-4537; d) K. Wang, H. Ling, Y. Bao, M. Yang, Y. Yang, M. Hussain, H. Wang, L. Zhang, L. Xie, M. Yi, *Adv. Mater.* **2018**, *30*, 1800595; e) Y. Katayama, M. Kalaj, K. S. Barcus, S. M. Cohen, *J. Am. Chem. Soc.* **2019**, *141*, 20000-20003; f) W. Zhu, G. Xiang, J. Shang, J. Guo, B. Motevalli, P. Durfee, J. O. Agola, E.

N. Coker, C. J. Brinker, *Adv. Funct. Mater.* **2018**, *28*, 1705274; g) T. Yun, H. Kim, A. Iqbal, Y. S. Cho, G. S. Lee, M. K. Kim, S. J. Kim, D. Kim, Y. Gogotsi, S. O. Kim, *Adv. Mater.* **2020**, *32*, 1906769; h) G. Han, S. Liu, Q. Yang, F. Zeng, W. Li, X. Mao, J. Xu, J. Zhu, *Polymer* **2022**, *259*, 125308.

[15] F. Liebig, R. M. Sarhan, M. Sander, W. Koopman, R. Schuetz, M. Bargheer, J. Koetz, *ACS Appl. Mater. Interfaces* **2017**, *9*, 20247-20253.

[16] T. Udayabhaskararao, T. Altantzis, L. Houben, M. Coronado-Puchau, J. Langer, R. Popovitz-Biro, L. M. Liz-Marzán, L. Vuković, P. Král, S. Bals, *Science* **2017**, *358*, 514-518.

[17] K. Volk, J. P. Fitzgerald, P. Ruckdeschel, M. Retsch, T. A. König, M. Karg, *Adv. Opt. Mater.* **2017**, *5*, 1600971.

[18] J. Miao, C. Chen, L. Meng, Y. Lin, *ACS Sensors* **2019**, *4*, 1279-1290.

[19] X. Li, J. F. Gilchrist, *Langmuir* **2016**, *32*, 1220-1226.

[20] R. Momper, H. Zhang, S. Chen, H. Halim, E. Johannes, S. Yordanov, D. Braga, B. Blülle, D. Doblas, T. Kraus, *Nano Lett.* **2020**, *20*, 4102-4110.

[21] S. H. Cho, K. M. Roccapriore, C. K. Dass, S. Ghosh, J. Choi, J. Noh, L. C. Reimnitz, S. Heo, K. Kim, K. Xie, *J. Chem. Phys.* **2020**, *152*, 014709.

[22] Y. Wu, S. Li, N. Gogotsi, T. Zhao, B. Fleury, C. R. Kagan, C. B. Murray, J. B. Baxter, *J. Phys. Chem. C* **2017**, *121*, 4146-4157.

[23] H. L. Nie, X. Dou, Z. Tang, H. D. Jang, J. Huang, *J. Am. Chem. Soc.* **2015**, *137*, 10683-10688.

[24] D. O. Schmidt, N. Raab, M. Noyong, V. Santhanam, R. Dittmann, U. Simon, *Nanomaterials* **2017**, *7*, 370.

[25] Q. Cao, S.-j. Han, G. S. Tulevski, Y. Zhu, D. D. Lu, W. Haensch, *Nat. Nanotechnol.* **2013**, *8*, 180-186.

[26] S. G. Booth, R. A. Dryfe, *J. Phys. Chem. C* **2015**, *119*, 23295-23309.

[27] X. Yu, M. S. Prévot, N. Guijarro, K. Sivula, *Nat. Commun.* **2015**, *6*, 7596.

[28] S. Lin, X. Lin, Y. Shang, S. Han, W. Hasi, L. Wang, *J. Phys. Chem. C* **2019**, *123*, 24714-24722.

[29] H. Pu, Z. Huang, F. Xu, D. W. Sun, *Food Chem.* **2021**, *343*, 128548.

[30] K. Whitham, D.-M. Smilgies, T. Hanrath, *Chem. Mater.* **2018**, *30*, 54-63.

[31] J. Hu, E. W. Spotte-Smith, B. Pan, R. J. Garcia, C. Colosqui, I. P. Herman, *J. Phys. Chem. C* **2020**, *124*, 23949-23963.

[32] Y. Jiang, R. Chakroun, P. Gu, A. H. Gröschel, T. P. Russell, *Angew. Chem.* **2020**, *132*, 12851-12855.

[33] C. Li, Y. Xu, X. Li, Z. Ye, C. Yao, Q. Chen, Y. Zhang, S. E. Bell, *Adv. Mater. Interfaces* **2020**, *7*, 2000391.

[34] J. Neilson, M. P. Avery, B. Derby, *ACS Appl. Mater. Interfaces* **2020**, *12*, 25125-25134.

[35] X. Li, X. Lin, X. Zhao, H. Wang, Y. Liu, S. Lin, L. Wang, S. Cong, *Appl. Surf. Sci.* **2020**, *518*, 146217.

[36] L. Song, N. Qiu, Y. Huang, Q. Cheng, Y. Yang, H. Lin, F. Su, T. Chen, *Adv. Opt. Mater.* **2020**, *8*, 1902082.

[37] K. Rajoua, L. Baklouti, F. Favier, *J. Phys. Chem. C* **2015**, *119*, 10130-10139.

[38]  K. R. Jinkins, S. M. Foradori, V. Saraswat, R. M. Jacobberger, J. H. Dwyer, P. Gopalan, A. Berson, M. S. Arnold, *Sci. Adv.* **2021**, *7*, eabh0640.

[39]  L. Liu, J. Han, L. Xu, J. Zhou, C. Zhao, S. Ding, H. Shi, M. Xiao, L. Ding, Z. Ma, *Science* **2020**, *368*, 850-856.

[40]  R. D. Deegan, O. Bakajin, T. F. Dupont, G. Huber, S. R. Nagel, T. A. Witten, *Nature* **1997**, *389*, 827-829.

[41]  H. Zargartalebi, S. H. Hejazi, A. Sanati-Nezhad, *Nat. Commun.* **2022**, *13*, 3085.

[42]  S. L. Young, J. E. Kellon, J. E. Hutchison, *J. Am. Chem. Soc.* **2016**, *138*, 13975-13984.

[43]  J. Lee, G. Bhak, J. H. Lee, W. Park, M. Lee, D. Lee, N. L. Jeon, D. H. Jeong, K. Char, S. R. Paik, *Angew. Chem. Int. Ed.* **2015**, *54*, 4571-4576.

[44]  N. N. Nam, T. L. Bui, S. J. Son, S. W. Joo, *Adv. Funct. Mater.* **2019**, *29*, 1809146.

[45]  H. Hu, M. Pauly, O. Felix, G. Decher, *Nanoscale* **2017**, *9*, 1307-1314.

[46]  D. Dong, L. W. Yap, D. M. Smilgies, K. J. Si, Q. Shi, W. Cheng, *Nanoscale* **2018**, *10*, 5065-5071.

[47]  V. Flauraud, M. Mastrangeli, G. D. Bernasconi, J. Butet, D. T. Alexander, E. Shahrabi, O. J. Martin, J. Brugger, *Nat. Nanotechnol.* **2017**, *12*, 73-80.

[48]  H. Zhu, J. F. O. Masson, C. G. Bazuin, *Langmuir* **2019**, *35*, 5114-5124.

[49]  A. Qdemat, E. Kentzinger, J. Buitenhuis, U. Rücker, M. Ganeva, T. Brückel, *RSC Adv.* **2020**, *10*, 18339-18347.

[50]  G. Wu, Y. Zhao, D. Ge, Y. Zhao, L. Yang, S. Yang, *Adv. Mater. Interfaces* **2021**, *8*, 2000681.

[51]  S. Khanna, P. Marathey, H. Chaliyawala, N. Rajaram, D. Roy, R. Banerjee, I. Mukhopadhyay, *Colloids Surf. A Physicochem. Eng. Asp.* **2018**, *553*, 520-527.

[52]  Q. Hu, Y. Abbas, H. Abbas, M. R. Park, T. S. Yoon, C. J. Kang, *Microelectron. Eng.* **2016**, *160*, 49-53.

[53]  T. A. Shastry, J. W. T. Seo, J. J. Lopez, H. N. Arnold, J. Z. Kelter, V. K. Sangwan, L. J. Lauhon, T. J. Marks, M. C. Hersam, *Small* **2013**, *9*, 45-51.

[54]  T. P. Bigioni, X.-M. Lin, T. T. Nguyen, E. I. Corwin, T. A. Witten, H. M. Jaeger, *Nat. Mater.* **2006**, *5*, 265-270.

[55]  a) Y. Wang, P. Kanjanaboos, S. P. McBride, E. Barry, X.-M. Lin, H. M. Jaeger, *Faraday Discuss.* **2015**, *181*, 325-338; b) L. Yi, W. Jiao, K. Wu, L. Qian, X. Yu, Q. Xia, K. Mao, S. Yuan, S. Wang, Y. Jiang, *Nano Res.* **2015**, *8*, 2978-2987; c) M. Meyns, S. Willing, H. Lehmann, C. Klinke, *ACS Nano* **2015**, *9*, 6077-6087; d) C.-P. Jen, T. G. Amstislavskaya, K.-F. Chen, Y. H. Chen, *Plos One* **2015**, *10*, e0126641; e) C. A. Rezende, J. Shan, L.-T. Lee, G. Zalczer, H. Tenhu, *J. Phys. Chem. B* **2009**, *113*, 9786-9794.

[56]  a) C. Tang, Z. Wang, W. Zhang, S. Zhu, N. Ming, G. Sun, P. Sheng, *Phys. Rev. B* **2009**, *80*, 165401; b) K. E. Mueggenburg, X. M. Lin, R. H. Goldsmith, H. M. Jaeger, *Nat. Mater.* **2007**, *6*, 656-660.

[57]  L. Wu, P. O. Jubert, D. Berman, W. Imaino, A. Nelson, H. Zhu, S. Zhang, S. Sun, *Nano Lett.* **2014**, *14*, 3395-3399.

[58]  Y. H. Lee, C. L. Lay, W. Shi, H. K. Lee, Y. Yang, S. Li, X. Y. Ling, *Nat. Commun.* **2018**, *9*, 2769.

[59]   Q. Zhao, H. Hilal, J. Kim, W. Park, M. Haddadnezhad, J. Lee, W. Park, J. W. Lee, S. Lee, I. Jung, *J. Am. Chem. Soc.* **2022**, *144*, 13285-13293.

[60]   C. F. Chen, S. D. Tzeng, H. Y. Chen, K. J. Lin, S. Gwo, *J. Am. Chem. Soc.* **2008**, *130*, 824-826.

[61]   J. M. Nam, J. W. Oh, H. Lee, Y. D. Suh, *Accounts Chem. Res.* **2016**, *49*, 2746-2755.

[62]   X. Lu, Y. Huang, B. Liu, L. Zhang, L. Song, J. Zhang, A. Zhang, T. Chen, *Chem. Mater.* **2018**, *30*, 1989-1997.

[63]   a) H. Liu, M. Siron, M. Gao, D. Lu, Y. Bekenstein, D. Zhang, L. Dou, A. P. Alivisatos, P. Yang, *Nano Res.* **2020**, *13*, 1453-1458; b) H. Hu, S. Wang, S. Wang, G. Liu, T. Cao, Y. Long, *Adv. Funct. Mater.* **2019**, *29*, 1902922.

[64]   D. Kim, W. K. Bae, S. H. Kim, D. C. Lee, *Nano Lett.* **2019**, *19*, 963-970.

[65]   S. H. Kang, W. S. Hwang, Z. Lin, S. H. Kwon, S. W. Hong, *Nano Lett.* **2015**, *15*, 7913-7920.

[66]   R. E. Smalley, M. S. Dresselhaus, G. Dresselhaus, P. Avouris, *Carbon Nanotubes: Synthesis, Structure, Properties and Applications*. **2003**.

[67]   Y. Joo, G. J. Brady, M. S. Arnold, P. Gopalan, *Langmuir* **2014**, *30*, 3460-3466.

[68]   K. R. Jinkins, S. M. Foradori, V. Saraswat, R. M. Jacobberger, J. H. Dwyer, P. Gopalan, A. Berson, M. S. Arnold, *Sci. Adv.* **2021**, *7*, eabh0640.

[69]   Q. Cao, S. j. Han, G. S. Tulevski, Y. Zhu, D. D. Lu, W. Haensch, *Nat. Nanotechnol.* **2013**, *8*, 180-186.

[70]   T. Dürkop, S. A. Getty, E. Cobas, M. Fuhrer, *Nano Lett.* **2004**, *4*, 35-39.

[71]   B. K. Sarker, S. Shekhar, S. I. Khondaker, *ACS Nano* **2011**, *5*, 6297-6305.

[72]   N. Rouhi, D. Jain, P. J. Burke, *ACS Nano* **2011**, *5*, 8471-8487.

[73]   M. Zheng, *Single-Walled Carbon Nanotubes: Preparation, Properties and Applications* **2019**, *375*, 129-164.

[74]   a) X. Tu, S. Manohar, A. Jagota, M. Zheng, *Nature* **2009**, *460*, 250-253; b) M. Zheng, A. Jagota, M. S. Strano, A. P. Santos, P. Barone, S. G. Chou, B. A. Diner, M. S. Dresselhaus, R. S. Mclean, G. B. Onoa, *Science* **2003**, *302*, 1545-1548; c) A. Nish, J.-Y. Hwang, J. Doig, R. J. Nicholas, *Nat. Nanotechnol.* **2007**, *2*, 640-646; d) H. W. Lee, Y. Yoon, S. Park, J. H. Oh, S. Hong, L. S. Liyanage, H. Wang, S. Morishita, N. Patil, Y. J. Park, *Nat. Commun.* **2011**, *2*, 541; e) H. Wang, Z. Bao, *Nano Today* **2015**, *10*, 737-758; f) C. Y. Khripin, J. A. Fagan, M. Zheng, *J. Am. Chem. Soc.* **2013**, *135*, 6822-6825; g) J. A. Fagan, C. Y. Khripin, C. A. Silvera Batista, J. R. Simpson, E. H. Hároz, A. R. Hight Walker, M. Zheng, *Adv. Mater.* **2014**, *26*, 2800-2804; h) H. Li, G. Gordeev, O. Garrity, N. A. Peyyety, P. B. Selvasundaram, S. Dehm, R. Krupke, S. Cambré, W. Wenseleers, S. Reich, *ACS Nano* **2019**, *14*, 948-963; i) H. Liu, D. Nishide, T. Tanaka, H. Kataura, *Nat. Commun.* **2011**, *2*, 309.

[75]   Q. Cao, *Nano Res.* **2021**, *14*, 3051-3069.

[76]   Y. Lin, Y. Cao, S. Ding, P. Zhang, L. Xu, C. Liu, Q. Hu, C. Jin, L.-M. Peng, Z. Zhang, *Nat. Electron.* **2023**, *6*, 506-515.

[77]   D. Zhong, C. Zhao, L. Liu, Z. Zhang, L. M. Peng, *Appl. Phys. Lett.* **2018**, *112*, 153109.

[78]    a) D. Zhong, Z. Zhang, L. Ding, J. Han, M. Xiao, J. Si, L. Xu, C. Qiu, L.-M. Peng, *Nat. Electron.* **2018**, *1*, 40-45; b) G. J. Brady, A. J. Way, N. S. Safron, H. T. Evensen, P. Gopalan, M. S. Arnold, *Sci. Adv.* **2016**, *2*, e1601240; c) Q. Cao, J. Tersoff, D. B. Farmer, Y. Zhu, S. J. Han, *Science* **2017**, *356*, 1369-1372; d) S. Yang, S. Ahmed, B. Arcot, R. Arghavani, P. Bai, S. Chambers, P. Charvat, R. Cotner, R. Gasser, T. Ghani, in *International Electron Devices Meeting 1998. Technical Digest (Cat. No. 98CH36217)*, IEEE, **1998**, 197-200; e) Y. Cao, G. J. Brady, H. Gui, C. Rutherglen, M. S. Arnold, C. Zhou, *ACS Nano* **2016**, *10*, 6782-6790; f) S. Tyagi, M. Alavi, R. Bigwood, T. Bramblett, J. Brandenburg, W. Chen, B. Crew, M. Hussein, P. Jacob, C. Kenyon, in *International Electron Devices Meeting 2000. Technical Digest. IEDM (Cat. No. 00CH37138)*, IEEE, **2000**, 567-570; g) M. Bohr, S. Ahmed, L. Brigham, R. Chau, R. Gasser, R. Green, W. Hargrove, E. Lee, R. Natter, S. Thompson, in *Proceedings of 1994 IEEE International Electron Devices Meeting*, IEEE, **1994**, 273-276.

[79]    L. Liu, L. Ding, D. Zhong, J. Han, S. Wang, Q. Meng, C. Qiu, X. Zhang, L.-M. Peng, Z. Zhang, *ACS Nano* **2019**, *13*, 2526-2535.

[80]    Y. Liu, S. Wang, H. Liu, L. M. Peng, *Nat. Commun.* **2017**, *8*, 15649.

[81]    S. Y. Ding, J. Yi, J.-F. Li, B. Ren, D.-Y. Wu, R. Panneerselvam, Z. Q. Tian, *Nat. Rev. Mater.* **2016**, *1*, 16021.

[82]    L. Tong, T. Zhu, Z. Liu, *Chem. Soc. Rev.* **2011**, *40*, 1296-1304.

[83]    X. Yu, H. Cai, W. Zhang, X. Li, N. Pan, Y. Luo, X. Wang, J. Hou, *ACS Nano* **2011**, *5*, 952-958.

[84]    K. Zhu, K. Yang, Y. Zhang, Z. Yang, Z. Qian, N. Li, L. Li, G. Jiang, T. Wang, S. Zong, *Small* **2022**, *18*, 2201508.

[85]    M. Kyung Oh, S. Park, S. K. Kim, S. H. Lim, *J. Comput. Theor. Nanosci.* **2010**, *7*, 1085-1094.

[86]    S. Si, W. Liang, Y. Sun, J. Huang, W. Ma, Z. Liang, Q. Bao, L. Jiang, *Adv. Funct. Mater.* **2016**, *26*, 8137-8145.

[87]    R. Esteban, A. G. Borisov, P. Nordlander, J. Aizpurua, *Nat. Commun.* **2012**, *3*, 825.

[88]    X. He, W. Gao, L. Xie, B. Li, Q. Zhang, S. Lei, J. M. Robinson, E. H. Hároz, S. K. Doorn, W. Wang, *Nat. Nanotechnol.* **2016**, *11*, 633-638.

[89]    W. Gao, X. Li, M. Bamba, J. Kono, *Nat. Photon.* **2018**, *12*, 362-367.

[90]    W. Gao, C. F. Doiron, X. Li, J. Kono, G. V. Naik, *ACS Photonics* **2019**, *6*, 1602-1609.

[91]    N. Komatsu, W. Gao, P. Chen, C. Guo, A. Babakhani, J. Kono, *Adv. Funct. Mater.* **2017**, *27*, 1606022.

[92]    J. Doumani, M. Lou, O. Dewey, N. Hong, J. Fan, A. Baydin, K. Zahn, Y. Yomogida, K. Yanagi, M. Pasquali, *Nat. Commun.* **2023**, *14*, 7380.

[93]    S. Goswami, A. J. Matula, S. P. Rath, S. Hedström, S. Saha, M. Annamalai, D. Sengupta, A. Patra, S. Ghosh, H. Jani, *Nat. Mater.* **2017**, *16*, 1216-1224.

[94]    a) P. A. Reissner, Y. Fedoryshyn, J. N. Tisserant, A. Stemmer, *Phys. Chem. Chem. Phys.* **2016**, *18*, 22783-22788; b) C. H. Kim, G. Bhak, J. Lee, S. Sung, S. Park, S. R. Paik, M. H. Yoon, *ACS Appl. Mater. Interfaces* **2016**, *8*, 11898-11903; c) T.

Zhang, D. Guérin, F. Alibart, D. Vuillaume, K. Lmimouni, S. Lenfant, A. Yassin, M. Oçafrain, P. Blanchard, J. Roncali, *J. Phys. Chem. C* **2017**, *121*, 10131-10139.

[95] A. Siria, M. L. Bocquet, L. Bocquet, *Nat. Rev. Chem.* **2017**, *1*, 0091.

[96] J. Xiao, M. Cong, M. Li, X. Zhang, Y. Zhang, X. Zhao, W. Lu, Z. Guo, X. Liang, G. Qing, *Adv. Funct. Mater.* **2023**, 2307996.

[97] a) M. Ebner, F. Marone, M. Stampanoni, V. Wood, *Science* **2013**, *342*, 716-720; b) C. Yuan, H. B. Wu, Y. Xie, X. W. Lou, *Angew. Chem. Int. Ed.* **2014**, *53*, 1488-1504.

[98] S. H. Lee, S. H. Yu, J. E. Lee, A. Jin, D. J. Lee, N. Lee, H. Jo, K. Shin, T. Y. Ahn, Y. W. Kim, *Nano Lett.* **2013**, *13*, 4249-4256.

[99] T. Li, B. Wang, J. Ning, W. Li, G. Guo, D. Han, B. Xue, J. Zou, G. Wu, Y. Yang, *Matter* **2019**, *1*, 976-987.

[100] Z. Liang, Q. Zhang, L. Jiang, G. Cao, *Energy Environ. Sci.* **2015**, *8*, 3442-3476.

[101] R. J. Ellingson, M. C. Beard, J. C. Johnson, P. Yu, O. I. Micic, A. J. Nozik, A. Shabaev, A. L. Efros, *Nano Lett.* **2005**, *5*, 865-871.

[102] B. J. Moon, S. Cho, K. S. Lee, S. Bae, S. Lee, J. Y. Hwang, B. Angadi, Y. Yi, M. Park, D. I. Son, *Adv. Energy Mater.* **2015**, *5*, 1401130.